\documentclass[aps,reprint, superscriptaddress, nofootinbib]{revtex4-1}

\usepackage{amsmath, amssymb, amsthm}
\usepackage{bm,bbm}
\usepackage{appendix}
\usepackage{graphicx}
\usepackage{nameref}
\usepackage{float}
\usepackage{hyperref}
\usepackage{booktabs}
\usepackage{dcolumn}
\usepackage[margin=2cm]{geometry}
\usepackage{array}
\usepackage{tabularx}
\usepackage{caption}
\usepackage{makecell}
\usepackage{ragged2e}
\usepackage{xcolor}
\usepackage{multirow}
\usepackage{tikz}
\usepackage{quantikz}
\usepackage{subcaption}
\usepackage{braket}
\usepackage{soul, color}

\theoremstyle{plain}
\newcommand{\be}{\begin{equation}}
\newcommand{\ee}{\end{equation}}
\newcommand{\bpm}{\begin{pmatrix}}
\newcommand{\epm}{\end{pmatrix}}

\begin{document}
\title{Practical insights on the effect of different encodings, ans\"atze and measurements in quantum and hybrid convolutional neural networks}

\author{Jes\'{u}s Lozano-Cruz\,\href{https://orcid.org/0009-0004-4029-2260}{\includegraphics[height=1.5ex]{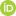}}{\ddag}}
\thanks{All authors contributed equally to this work.}
\affiliation{Quantum Lab, CTIC Centro Tecnol\'ogico, Parque Cient\'ifico Tecnol\'ogico de Gij\'on, Asturias, Spain}

\author{Albert Nieto-Morales\,\href{https://orcid.org/0009-0007-5060-6157}{\includegraphics[height=1.5ex]{Figures/orcid.png}}{\ddag}}\thanks{All authors contributed equally to this work.}
\affiliation{Quantum Lab, CTIC Centro Tecnol\'ogico, Parque Cient\'ifico Tecnol\'ogico de Gij\'on, Asturias, Spain}
\affiliation{Universidad Internacional de la Rioja (UNIR), Logro\~{n}o, Spain}

\author{Oriol Ball\'o-Gimbernat\,\href{https://orcid.org/0009-0000-0134-3206}{\includegraphics[height=1.5ex]{Figures/orcid.png}}}\thanks{All authors contributed equally to this work.}
\affiliation{Eurecat, Centre Tecnol\`ogic de Catalunya, Multimedia Technologies,
Barcelona, Spain}
\affiliation{Centre de Visi\'o per Computador (CVC), Barcelona, Spain.}
\affiliation{Universitat Aut\`onoma de Barcelona (UAB),  Barcelona, Spain.}

\author{Adan Garriga\,\href{https://orcid.org/0000-0003-2681-4005}{\includegraphics[height=1.5ex]{Figures/orcid.png}}}\thanks{All authors contributed equally to this work.}
\affiliation{Eurecat, Centre Tecnol\`ogic de Catalunya, Multimedia Technologies,
Barcelona, Spain}

\author{Ant\'on Rodr\'iguez-Otero}\thanks{All authors contributed equally to this work.}
\affiliation{Fsas International Quantum Center (Fujitsu), Santiago de Compostela, Spain.}

\author{Alejandro Borrallo-Rentero}\thanks{All authors contributed equally to this work.}
\affiliation{Fsas International Quantum Center (Fujitsu), Santiago de Compostela, Spain.}

\date{\today}


\renewcommand\thefootnote{\ddag} 
\footnotetext{Corresponding authors: jesus.lozano@fundacionctic.org and \\ albert.nieto@unir.net} 


\begin{abstract}

This study investigates the design choices of parameterized quantum circuits (PQCs) within quantum and hybrid convolutional neural network (HQNN and QCNN) architectures applied to the task of satellite image classification using the EuroSAT dataset. We systematically evaluate the performance implications of data encoding techniques, variational ans\"atze, and measurement in $\approx500$ distinct model configurations. Our analysis reveals a clear hierarchy of influence on model performance. For hybrid architectures, which were benchmarked against their direct classical equivalents (e.g., the same architecture with the PQCs removed), the data encoding strategy is the dominant factor, with validation accuracy varying over $30\%$ for distinct embeddings. In contrast, the selection of variational ans\"atze and measurement basis had a comparatively marginal effect, with validation accuracy variations remaining below $5\%$. For purely quantum models, restricted to amplitude encoding, performance was most dependent on the measurement protocol and the data-to-amplitude mapping. The measurement strategy varied the validation accuracy by up to $30\%$ and the encoding mapping by around $8$ percentage points.

\textbf{Keywords:} quantum machine learning, parameterized quantum circuit, image classification, quantum neural network, ansatz

\end{abstract}

\maketitle

\section{Introduction}
In recent years, quantum machine learning (QML) has emerged as an interdisciplinary field at the intersection of quantum computing and machine learning \cite{Schuld2014, Wittek2014, Biamonte2017, Schuld2018}. Its objective is to leverage phenomena in quantum mechanics to tackle problems from machine learning. A central component of many contemporary QML models is the parameterized quantum circuit (PQC), a flexible framework for creating trainable quantum circuits that can be used independently or integrated as hybrid components within classical neural networks (NNs).

This work investigates the use of PQCs for the task of satellite image classification using the EuroSAT dataset \cite{Helber2019}. These PQCs serve as trainable quantum components that can be integrated into NNs or used by themselves. While PQCs offer a flexible paradigm for hybrid and purely QML models, their efficacy is highly sensitive to a multitude of design choices, with the most important being the encoding technique, the ans\"atze and the measurement strategy.

We study two different hybrid architecture networks and a purely quantum one, all inspired by convolutional neural networks (CNN). For the hybrid ones, we have a ``Quantum-First'' model, in which PQCs are used as convolutions to extract features, and a ``Quantum-Later'' model \cite{Senokosov2024}, in which PQCs are used between the fully connected layers to process the extracted features \cite{Senokosov2024,Hur2022,Elhag2024}. For the quantum architecture, we use three variations of PQCs as standalone components, one of which is inspired by the QCNN, introduced in \cite{Cong2019}.

While the work of \cite{Bowles2024} provides a foundational benchmark focusing on the scaling of hyperparameters and the theoretical implications of entanglement, our work pivots to the structural composability of QML architectures. We argue that optimizing hyperparameters is complementary to selecting the correct combination of fundamental blocks. Therefore, rather than performing a hyperparameter grid search, we systematically evaluate the tripartite interaction between encoding, ans\"atze and measurement. Crucially, we analyze how this interplay varies across a spectrum of architectural paradigms, ranging from hybrid integrations (``Quantum-First'' and ``Quantum-Later'') to standalone QCNNs. This comprehensive scope allows us to distinguish between insights applicable generally to PQCs and those specific to structured quantum architectures, an angle not covered in \cite{Bowles2024}.

The primary objective of this work is not necessarily to achieve state-of-the-art performance, or claim quantum utility, but rather to find practical insights into how core PQC design elements influence model behavior and efficacy. The findings are intended to aid researchers and practitioners in designing QML models.

All our experiments were done through simulations using the \textit{CTIC Quantum Testbed (QUTE)} platform \cite{qute}. We used PennyLane \cite{bergholmPennyLaneAutomaticDifferentiation2022} to implement the PQCs, both in the purely quantum and hybrid models, and PyTorch \cite{paszke2019pytorchimperativestylehighperformance} for the classical layers.

In the following \hyperref[sec:methods]{Materials and Methods Section}, we detail the dataset utilized for our experiments and explain the development process of the parameterized quantum circuits underlying our methodology. The \hyperref[sec:models]{Models Section} presents the architectural designs investigated, along with the reasoning guiding their formulation. Empirical outcomes demonstrating model performance across diverse scenarios are summarized in the \hyperref[sec:results]{Results Section}. Lastly, the \hyperref[sec:evaluation]{Evaluation Section} critically examines these findings, emphasizing key metrics and offering a balanced perspective on the efficacy and constraints of our approach.

\section{Materials and methods}\label{sec:methods}
\subsection{Image classification}\label{ImageClassification}
\begin{figure}[h]
 \includegraphics[width=1\columnwidth]{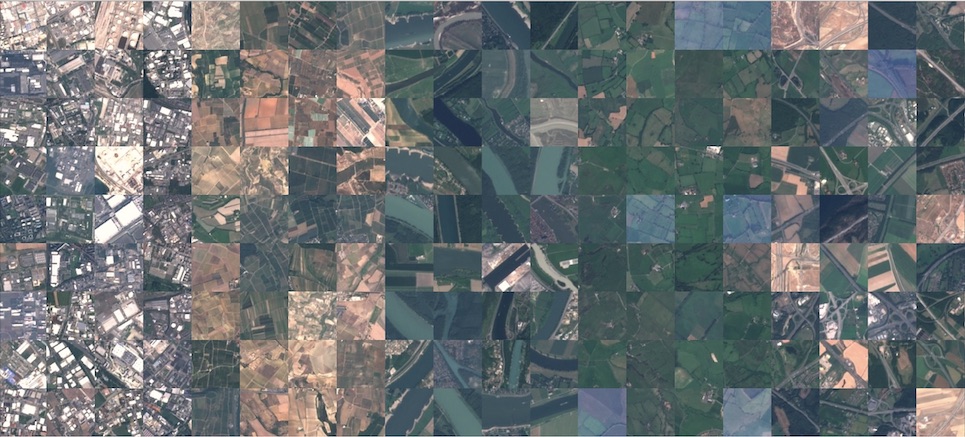}
 \caption{Samples from the EuroSAT dataset introduced in \cite{Helber2019}. Each sample is a $64\times64\times3$ RGB image obtained from the Sentinel-2 satellite.}
    \label{EUROSAT}
\end{figure}

Our focus will be on classifying satellite images from the EuroSAT dataset \cite{Helber2019}; examples can be seen in \hyperref[EUROSAT]{FIG. \ref*{EUROSAT}}. EuroSAT was selected as our target due to its high intra-class variance and its standing as a standard, challenging dataset for evaluating remote sensing architectures. The dataset consists of $27,000$ geo-referenced images, provided by the Copernicus Earth observation program \cite{Copernicus} using the Sentinel-2 satellite, and comes in two versions: RGB and all bands. The first consists of only RGB channels, and the other uses all the bands of the spectrum as collected from Sentinel-2. Each image channel has a resolution of $64\times64$ pixels.

The classification task involves distinguishing between 10 distinct land-cover classes: annual crop, forest, herbaceous vegetation, highway, industrial, pasture, permanent crop, residential, river, and sea. The dataset is balanced, with $2000$ to $3000$ images per class, posing a significant challenge for models to identify specific textural and spatial patterns across diverse geographical regions.

In our experiments, we used the RGB version and downscaled them to $32\times32$ and $16\times16$. This downsampling represents a practical trade-off: satellite imagery at these scales preserves the essential spatial and spectral features required for land-cover classification. This is evidenced by the structural similarity index remaining above $0.9$ for the first case and over $0.78$ for the latter. Furthermore, this dimensionality reduction is vital for research in the current NISQ era; it ensures that the architectures remain computationally tractable for state-of-the-art simulators, allowing for the rigorous evaluation of hybrid models while the field transitions toward fault-tolerant hardware. Notably, the $32\times32$ version was only studied for the hybrid models because of computation constraints. The $16\times 16$ version was studied for both hybrid and purely quantum models.

We selected this dataset as a continuation of the work conducted within the ARQA Cervera Technology Transfer Network \cite{ARQA}. Nevertheless, identical experiments can be replicated on alternative datasets, provided they satisfy the previously established size constraints.

\subsection{Parameterized quantum circuits}\label{PQC}
\begin{figure}[h]
    \centering
    \resizebox{\linewidth}{!}{
    \begin{quantikz}
    \slice{Classical input} &\gate[5]{{Embedding}}&\gate[5]{{Ansatz}}& \meter{} & \slice{Classical output} &\\
    &&&\meter{}&&\\
    \vdots \\
    &&&\meter{}&&\\
    &&&\meter{}&&
    \end{quantikz}
    }
    \caption{Prototypical quantum circuit scheme.}
    \label{quantumlayer}
\end{figure}
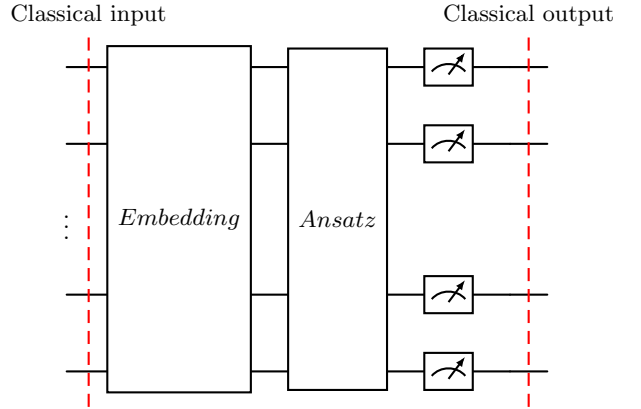

PQCs are the cornerstone of our approach. We integrate them as layers consisting of one or more circuits in the hybrid models and use them by themselves in the purely quantum ones. In general, PQCs consist of three key components: encoding, used to embed the information in a quantum state; variational ansatz, used to manipulate the encoded information; and measurement, used to extract information from the quantum state. \hyperref[quantumlayer]{FIG. \ref*{quantumlayer}} shows the prototypical example of a PQC.

\subsubsection*{Data encoding}
As previously mentioned, the first stage in a PQC is the data encoding; this step serves as a mapping from data to a quantum state. Given that there is not a one-size-fits-all solution for the encoding step, we chose to include the following ones in our analysis:

\begin{itemize}
    \item \textbf{Angle encoding} \cite{Schuld2018, Schuld2014b, Cao2017}:
    This is a prevalent and relatively straightforward method where \(N\)
    classical data features, \(x_1, x_2, \ldots, x_N\), are encoded
    into the rotation angles of \(N\) individual qubits. Each feature \(x_j\) is used as an angle parameter for a single-qubit
    rotation gate, such as \(R_X(\theta_j)\), \(R_Y(\theta_j)\), or
    \(R_Z(\theta_j)\), applied to a specific qubit. Notably, using \(R_Z(\theta_j)\) as an encoding gate acting on an initial state from the computational basis only adds an irrelevant global phase. The cost of this embedding is $\mathcal{O}(1)$.

    \item \textbf{Amplitude encoding} \cite{Schuld2018}:
        This technique embeds a classical data vector of \(2^n\)
        normalized real numbers, \(\{\alpha_i\}\), directly into the
        amplitudes of an \(n\)-qubit quantum state vector \(|\Psi\rangle\).
        The state is represented as
        $$ |\Psi\rangle = \sum_{i=0}^{2^n-1} \alpha_i |i\rangle $$
        where \(|i\rangle\) are the computational basis states, and the
        amplitudes must satisfy the normalization condition
        \(\sum_{i=0}^{2^n-1} |\alpha_i|^2 = 1\). The state preparation process required to achieve an
        arbitrary state \(|\Psi\rangle\) can be computationally
        expensive, potentially demanding an exponentially deep circuit \cite{knill1995approximation}.
    \item \textbf{Instantaneous quantum polynomial-time (IQP) encoding} \cite{Hav2019}:
        This encoding extends the concept of angle encoding by incorporating
        entangling operations. \(N\) classical features  are typically
        encoded into \(N\) qubits using single-qubit rotations, akin to
        angle encoding. These rotations are then interspersed with diagonal
        two-qubit gates, commonly \(R_{ZZ}(\phi_{jk})\) gates, where
        \(\phi_{jk}\) can be a function of individual features or products
        of features (e.g., \(x_j x_k\)). This structure allows for the
        generation of more complex quantum states that can potentially capture
        correlations between input features. The number of layers, $L$, is a tunable parameter. The scaling of this embedding is dominated by the two-qubit operations which in principle involve all pairs of qubits, thus making the scaling $\mathcal{O}(LN^2)$.
    \item \textbf{Quantum approximate optimization algorithm (QAOA) inspired encoding} \cite{Lloyd2020}:
        This encoding strategy draws inspiration from the structure of the
        QAOA. It transforms \(N\) classical features into parameters that
        define a QAOA-like ansatz applied to \(N\) qubits. The features typically determine the angles for rotation gates within layers corresponding to a problem Hamiltonian and a mixer Hamiltonian. The number of layers $L$ is a tunable parameter, and there are 3 variants: $X$, $Y$ and $Z$. The scaling of this embedding is dominated by the trainable two-qubit ZZ interactions, which connect nearest neighbors; hence the scaling is $\mathcal{O}(LN)$.
    \item \textbf{Ring encoding} \cite{Venturelli2023}:
        This method combines angle encoding for \(N\) features on \(N\)
        qubits with a specific entangling topology. Following initial
        single-qubit rotations that embed the classical data, two-qubit
        entangling gates are applied between
        adjacent qubits arranged in a ring structure. Specifically, qubit
        \(i\) is entangled with qubit \((i+1) \mod N\); thus, $N$ two-qubit entangling gates are applied. The number of layers, $L$, is a tunable parameter. Since consecutive entangling gates share one qubit, the cost of ring encoding is $\mathcal{O}(N L)$.

    \item \textbf{Waterfall encoding} \cite{Venturelli2023}:
    Similar to Ring encoding, Waterfall encoding also utilizes angle
    encoding for \(N\) features on \(N\) qubits, followed by a
    structured application of entangling gates. In this scheme, the
    entangling gates are typically arranged in a
    ``waterfall'' pattern. For example, qubit \(i\) might
    control an operation on qubit \(i+1\), which in turn might control
    an operation on qubit \(i+2\), and so forth. In each layer, there are $N(N-1)/2$ entangling gates; hence, worst case scaling is $\mathcal{O}(L N^2)$. 
\end{itemize} 

\subsubsection*{Variational ansatz}
The variational circuit constitutes the second part of a PQC. It consists of a set of quantum gates with trainable parameters, akin to neurons in a NN. Its purpose is to manipulate the quantum state generated by the previous step with the goal of turning it into a state whose measurement, in a given basis, solves the task at hand. We chose to experiment with the following ones:

\begin{itemize}
    \item \textbf{No entanglement}: This circuit uses randomness in the assignment of quantum gates, applying either an \emph{RZ} gate or the identity operation to each qubit in a stochastic manner. Due to the potential presence of trivial operations (identity), the resulting circuit typically exhibits a lower computational depth compared to more densely parameterized or entangled circuits.
    \item \textbf{Full entanglement} \cite{Schuld2020}: This circuit is implemented using the \texttt{StronglyEntanglingLayers} module from PennyLane \cite{bergholmPennyLaneAutomaticDifferentiation2022}. It allows for a configurable number of layers, each consisting of a sequence of single-qubit rotations followed by entangling operations. This structure enables flexible depth and entanglement patterns suitable for various variational quantum algorithms.
    \begin{figure}[htbp]
        \centering 
        \includegraphics[width=0.42\textwidth]{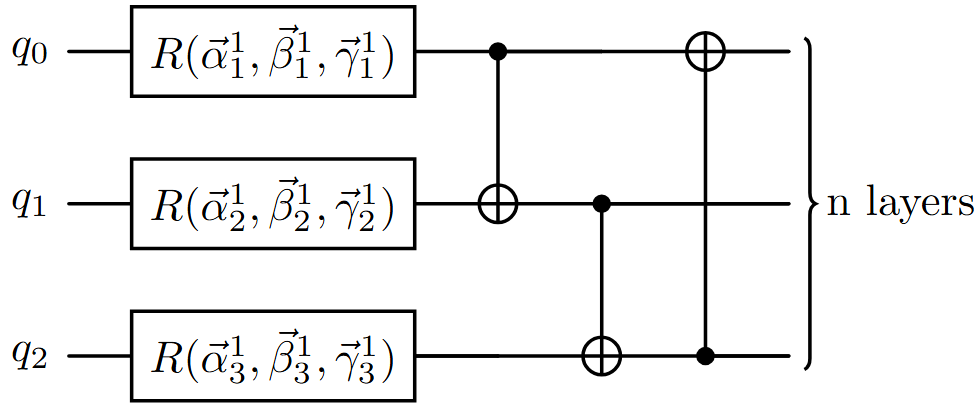} 
        \caption{Visualization of Full entanglement ansatz.}
        \label{fig:strong} 
    \end{figure}

    \item \textbf{Ring circuit} \cite{Venturelli2023}: It performs single-qubit rotations using \emph{RY} and \emph{RZ} gates, followed by the application of controlled-Z (\emph{CZ}) gates arranged in a ring topology. The resulting structure resembles the ring embedding, but with entanglement introduced through \emph{CZ} operations instead of \emph{CNOT}. 
    \begin{figure}[htbp]
        \centering 
        \includegraphics[width=0.42\textwidth]{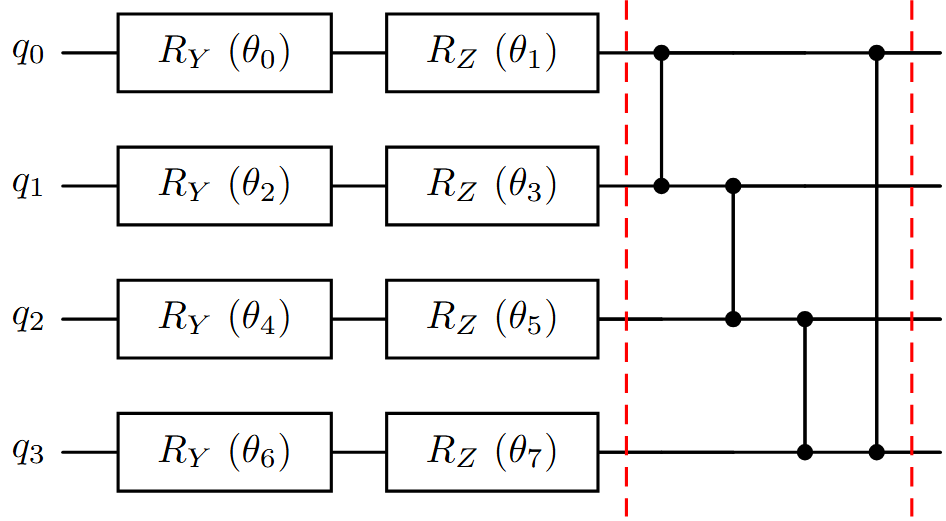} 
        \caption{Visualization of Ring ansatz.}
        \label{fig:ring} 
    \end{figure}
    
    \item \textbf{NQ circuit} \cite{Hur2024}: The circuit consists of multiple layers, each combining parameterized single-qubit rotations, \emph{RY} and \emph{RZ}, followed by entangling gates (\emph{CNOT}) in a linear nearest-neighbor topology.
    \begin{figure}[htbp]
        \centering 
        \includegraphics[width=0.44\textwidth]{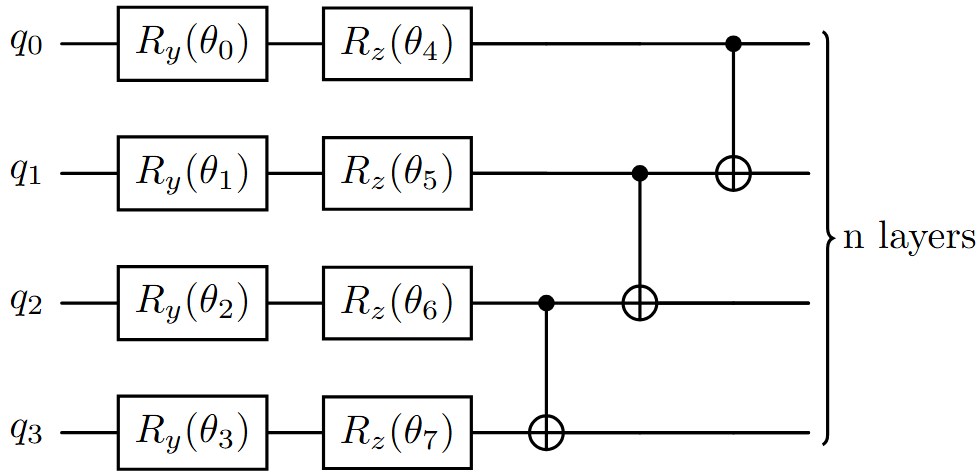} 
        \caption{Visualization of NQ ansatz.}
        \label{fig:NQE} 
    \end{figure}
    
    \item \textbf{QCNN circuit} \cite{Cong2019}: circuit structure inspired by classical CNNs in which convolutional and pooling operations are subsequently applied in each layer (see \hyperref[qcnnfc]{FIG. \ref*{qcnnfc}}). In our implementation, when the number of qubits at a layer is odd, one is excluded from the pooling, although it is still considered in the convolutional stages. At the end of the circuit, the qubits that were discarded are recovered for the fully connected layer. Given that the number of free parameters in an arbitrary unitary scales as $4^N-1$, we substituted this last operation by the \emph{Simplified Two Design} \cite{cerezo2021cost}.
\end{itemize}

\subsubsection*{Measurement}
The final step of a PQC is the measurement; at this stage, either all or only a fraction of the qubits are measured to extract classical information. 
This information can then be used to update parameters, feed the next layer in a hybrid network, or serve as the final output. In pure quantum models, we do not need to measure until the end of the network. We tested with 3 measurements in the hybrid case.

\begin{itemize}
    \item \textbf{Z-basis}: It is also called the computational basis. This is the most commonly used measurement in quantum circuits because it collapses a qubit into $|0\rangle$ or $|1\rangle$.
    \item \textbf{X-basis}: In this case, a Hadamard gate (\emph{H}) is applied before measuring in the Z-basis. It is often used in entangled states.
    \item \textbf{Y-basis}: Now, the inverse of a phase gate ($S^\dagger$) is applied and then measures in the X-basis. This is typically employed in phase-sensitive protocols.
    
\end{itemize}

In the fully quantum models, in order to have more flexibility when it comes to the amount of classical information we can extract out of a circuit, we also measured multiqubit observables:

\begin{itemize}
    \item \textbf{Histogram}: We measure each qubit on the Z basis and count the frequency of each of the possible $2^N$ outcomes, for $N$ qubits. Each state is associated with a class; hence, the model in this case will try to learn to link classes with corresponding states. This is in essence a Quantum Circuit Born Machine \cite{liu2018differentiable} in which we set the most frequent sample to be the predicted class label.
    \item \textbf{Pauli string (Paulis)}: We estimate the expectation value of Pauli strings of $N$ qubits, i.e., observables of the form $O = \otimes_{i=0}^N P_i$, where $P_i$ is a Pauli gate acting on the $i-th$ qubit. Here, a class is identified with an operator such that the model is trained to maximize the expectation value of that observable when images from the correct class are fed into the model.
\end{itemize}

Recent studies have also pointed out the relevance of the type of measurement considered in the models \cite{chen2025learningprogramquantummeasurements}. 

\section{Models}\label{sec:models}

For an overview of the models, see \hyperref[table1]{TABLE \ref{table1}}. This table outlines the distinct components integrated into each model, including data encodings, ansätze, and measurement strategies.

\subsection{Hybrid models I: HQNN-Parallel}
The \emph{HQNN-Parallel} was first introduced in \cite{Senokosov2024} (shown in \hyperref[hqnn]{FIG. \ref*{hqnn}}). This architecture introduces a NN layer based on multiple PQCs. From this point onward, we will call this a quantum linear layer due to its similarity with a classical linear layer. 

A quantum linear layer operates similarly to its classical counterpart. It takes in an input vector of size $V$ and outputs another of size $W$. However, unlike the classical case, here $W\leq V$ given that at most we do a single measurement on all qubits. In the classical version, the linear layer is defined by the number of input and output features. The quantum analog is characterized by the number of input features, the number of qubits allocated per circuit (the same number for all), and the encoding used, which will determine the proportion between input and output features. 

\begin{figure}[htbp]
    \centering 
    \includegraphics[width=0.45\textwidth]{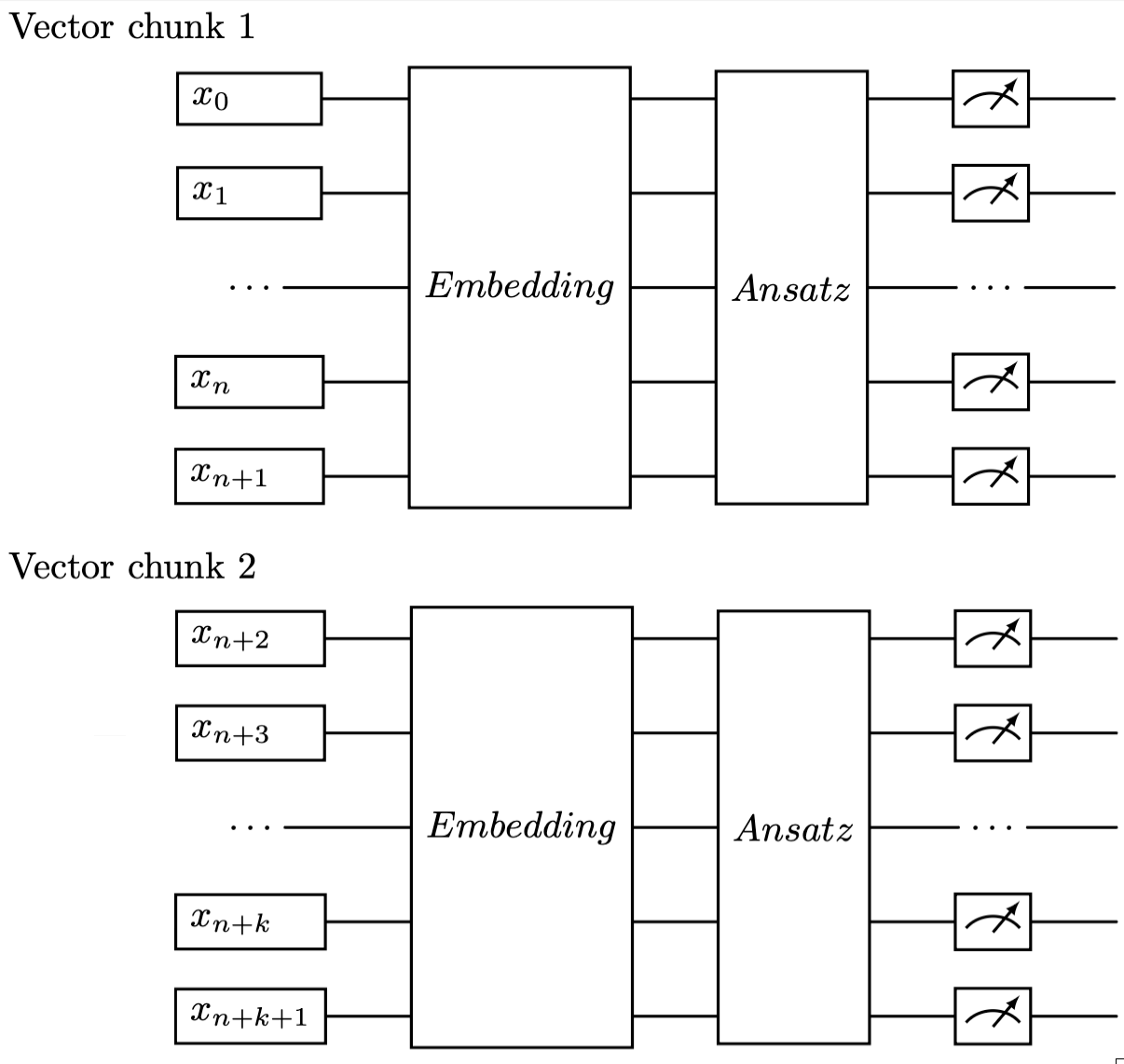} 
    \caption{Depiction of the quantum linear layer for processing vector $\mathbf{x}$ with two circuits. The circuits independently process their corresponding vector chunk and can be run in parallel.}
    \label{fig:quantum-linear-layer} 
\end{figure}

In the simplest case, we have an input vector of length $L$, we allocate $L/2$ qubits per circuit, and we use an encoding that maps a single feature to a single qubit, like angle encoding. In this case, the quantum linear layer will consist of two PQCs of $L/2$ qubits each, and the output will be a vector of length $L$, given that we measure all qubits on a given basis. This scheme is shown in \hyperref[fig:quantum-linear-layer]{FIG. \ref*{fig:quantum-linear-layer}}. For instance, if we use amplitude embedding, then each $N$ qubit circuit will process $2^N$ features, and the output will be of size $N\times \text{the number of circuits}$.

This architecture scales really well with the available quantum resources, given that you can tune the number of qubits allocated per circuit. The higher the qubit count per circuit, the more correlations between features will be accessible, and better performance is expected. Moreover, given that all circuits are independent, they can be executed in parallel.

To ensure a consistent dimensionality for the feature across the different input scales, we dynamically adjust the depth of the feature extractor. For $32\times32\times3$ inputs, we employ a four-layer convolutional block that reduces the spatial dimensions to $4\times4$ with $48$ channels. For $16\times16$ inputs, we use a two-layer block, resulting in an $8\times8$ feature map with $12$ channels. In both configurations, the resulting flattened feature vector remains constant at $768$ elements. This architectural consistency allows us to utilize a standardized quantum linear layer across all experiments, consisting of $96$ circuits for cases using one-to-one encodings and $3$ circuits for those utilizing amplitude encoding. 

\subsection{Hybrid models II: HQNN-Quanv}
\label{sec:hqnn_quanv}

\begin{figure*}[!htbp]
    \centering
    \begin{subfigure}{0.95\textwidth}
        \centering
        \includegraphics[width=\linewidth]{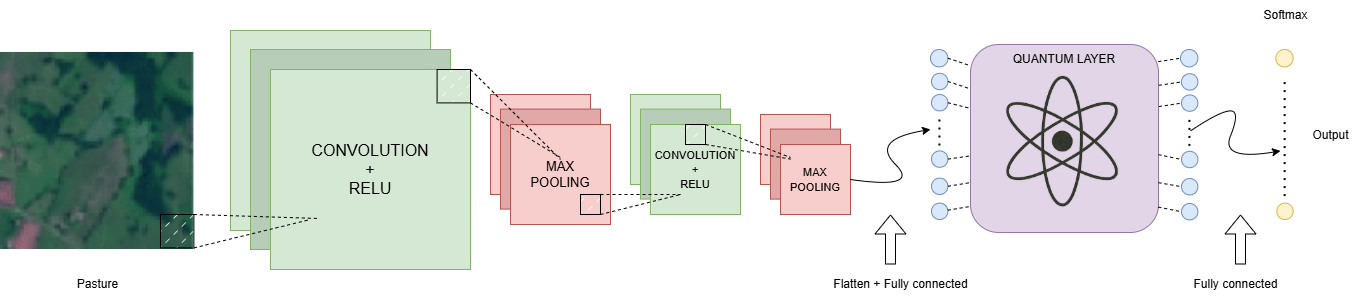}
        \subcaption{ HQNN-Parallel model from \cite{Senokosov2024}.}
        \vspace{0.05\textwidth} 
        \label{hqnn}
    \end{subfigure}
    \vspace{0.05\textwidth} 
    \begin{subfigure}{0.85\textwidth}\label{quanv}
        \centering
        \includegraphics[width=\linewidth]{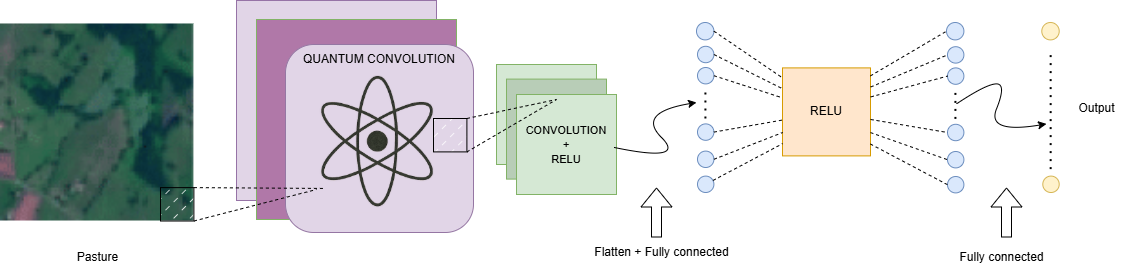} 
        \subcaption{ HQNN-Quanv model inspired by \cite{Simonyan2014}.}
        \label{flex}
    \end{subfigure}
    \vspace{0.05\textwidth} 
    \begin{subfigure}{0.90\textwidth}
        \centering
        \includegraphics[width=\linewidth]{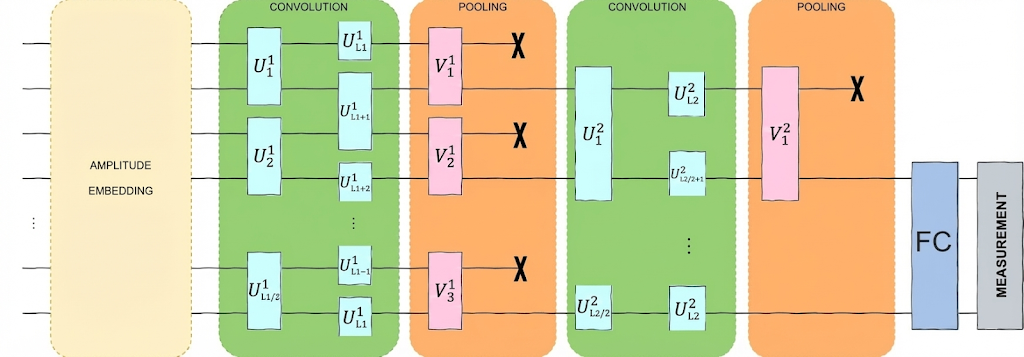} 
        \subcaption{ QCNN model from \cite{Cong2019}.}
        \label{qcnnfc}
    \end{subfigure}
    \vspace{0.05\textwidth} 
    \begin{subfigure}{0.9\textwidth}\label{double}
        \centering
        \begin{subfigure}{0.48\textwidth}
            \centering
            \includegraphics[width=\linewidth]{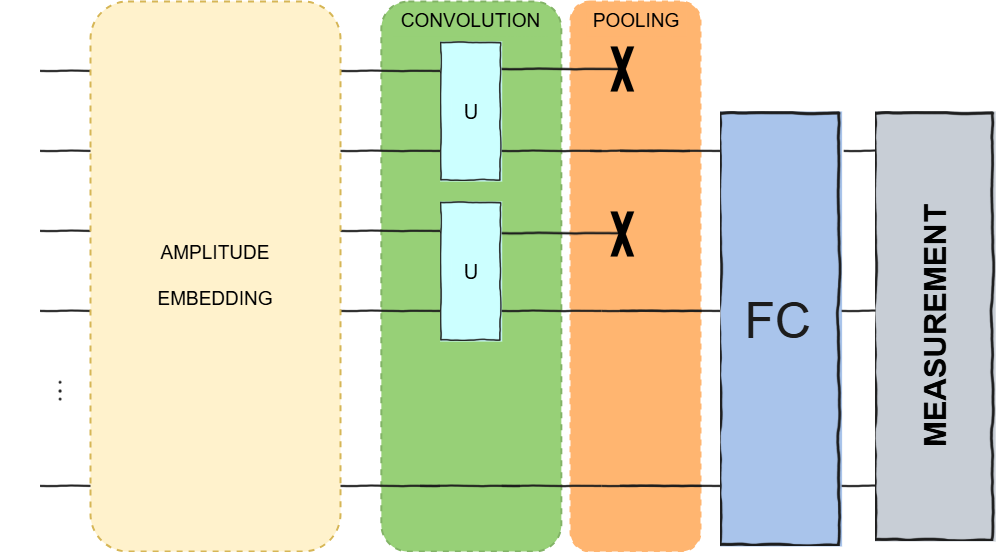}
            \subcaption{Two kernels.}
            \label{twokernel}
        \end{subfigure}
        \hfill
        \begin{subfigure}{0.48\textwidth}
            \centering
            \includegraphics[width=\linewidth, height=4.25cm]{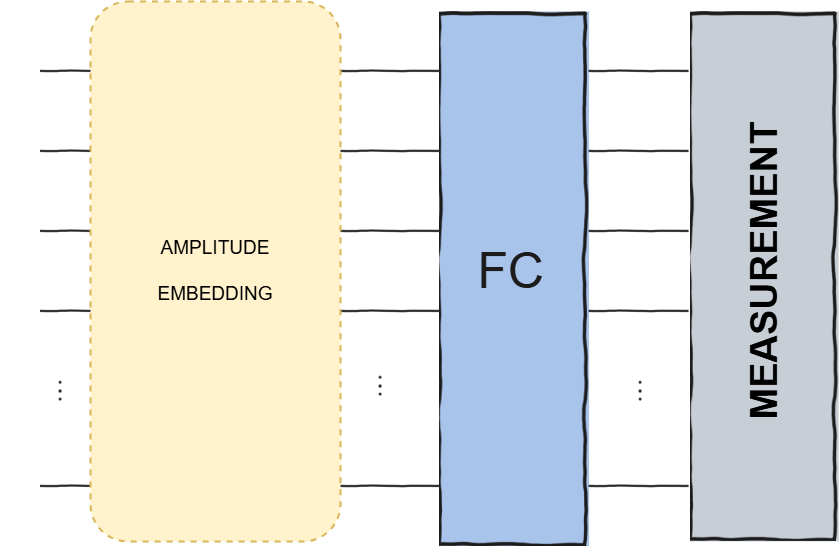}
            \subcaption{FC model.}
            \label{fcmodel}
        \end{subfigure}
    \end{subfigure}
    \caption{Scheme of every model used. The three purely quantum models, \hyperref[qcnnfc]{c)}, \hyperref[twokernel]{d)} and \hyperref[fcmodel]{e)}, are particular cases of the generic PQC structure in Fig. \ref{quantumlayer}. Quantum circuits were created using draw.io, an open-source library \cite{drawio}.}
    \label{models}
\end{figure*}

The \emph{HQNN-Quanv} model follows the high-level blueprint of Senokosov \emph{et al.}~\cite{Senokosov2024} but embeds it in the depth hierarchy of VGG16~\cite{Simonyan2014}, replacing the very first classical convolution with a bank of small, parameter-tied quantum circuits (see \hyperref[flex]{FIG.~\ref*{flex}}). Every input image is split channel-wise into a dense grid of overlapping patches of size $\text{qks}\times\text{qks}$ (we study quantum-kernel size $\text{(qks)}\!=\!2,3$ with stride~$1$).  Each patch is flattened, normalized, and loaded into a register of $Q=\text{qks}^{2}$ qubits with one of the encodings introduced in Sec.~\ref{PQC}.  Because the same circuit parameters are reused for every spatial position and color channel, the layer acts as a genuine quantum analogue of weight-tied convolution.

Inside each register, we apply two identical blocks of a variational ansatz. These blocks are deep enough to mix local features but shallow enough to keep simulation tractable; their parameters, shared across all patches, are the only quantum learnable weights.  Upon exit every qubit is measured once in the $X$, $Y$, or $Z$ basis, yielding $Q$ scalars per patch.  Aggregating those scalars across the full grid produces $C \times Q$ feature maps of size $H'\!\times\!W'$, where $H'$ and $W'$ are the patch indices.  Here $C$ is the number of color channels that are processed independently ($C=3$ for RGB images).  A fixed \texttt{ReLU} is applied, and the tensor is passed to two classical $3\times3$ convolutions, global average pooling, and a pair of fully connected layers that emit softmax probabilities for the EuroSAT classes.

The learnable weights split into two groups: a small set $P_{\text{q}}$ of rotation angles inside one patch circuit, and the usual classical weights $P_{\text{c}}$ that follow. Because the \emph{same} circuit is reused for every patch, $P_{\text{q}}$ stays fixed when we move from $32\times32$ to $16\times16$ images or change the stride, while $P_{\text{c}}$ matches the classical head already described for HQNN-Parallel. In practice $P_{\text{q}}$ is only a few hundred parameters, roughly 10–100 times fewer than the weights in a classical first convolution that produces the same number of channels. Yet, it still injects a strong non-linearity at every pixel location.

\subsection{Pure quantum models I: QCNN}\label{pqm}
QCNNs were first introduced in \cite{Cong2019}. This model performs both the feature extraction and the classification with concatenated quantum operations, so the full model is represented as a unique quantum circuit (see \hyperref[qcnnfc]{FIG. \ref*{qcnnfc}}). 

It draws inspiration from classical CNN and proposes doing convolutions via 2-qubit gates acting on adjacent qubits; followed by a pooling operation, which reduces the dimensionality by tracing out a subset of the qubits. It had great success due to its remarkable performance given its trainability \cite{Hur2022,simen2024digital}. 

In this study, we did not want to consider an initial classical stage of dimensionality reduction, since the goal was to check how far we could go with only quantum components. Given limited computational resources, we did not want to exceed the range of 10 to 15 qubits; thus, we used amplitude embedding to be able to fit the images in the available number of qubits. 

However, if we store the pixels in the amplitudes of the wavefunction, then the analogy of performing a convolution by 2-local gates needs to be looked at again. It turns out that if you track the way in which pixels are mixed in this scenario, you see that pixels from the same vicinity are indeed being mixed (\hyperref[appendix:SEQNN]{Appendix \ref*{appendix:SEQNN}}), just like in a convolution. Moreover, by choosing the order in which the amplitudes are fed into the wavefunction, different kernel shapes can be implemented. In this work, we selected several orderings and checked their impact on the QCNN and a simplified model where the effect of the mixing is isolated: the Shaped Embedding Quantum Neural Network (SEQNN). 

\subsection{Pure quantum models II: SEQNN}\label{pqm2}
To test the intuition that the order in which pixels are introduced into the amplitude of the state vector affects the way in which pixels are mixed, we built two simplified models. 

\subsubsection*{FC}
This model is used as a baseline; it consists of just a block with the structure described by the \emph{Simplified Two Design} ansatz, which we used as a fully connected layer in the QCNN architecture.

\subsubsection*{Two kernels}
Consists of three parts (check \hyperref[twokernel]{FIG. \ref*{twokernel}}):
\begin{enumerate}
    \item Two arbitrary unitary operations acting as convolutional filters: one operating on the first two qubits and the second one on the following pair of qubits.
    \item A pooling operation: one of the qubits on which the arbitrary unitary operations are acting is traced out, that is, two in total.
    \item A fully connected layer on all remaining qubits.
\end{enumerate}

\subsection{Classical Baselines}
To strictly isolate the performance impact of the PQCs, we benchmarked each hybrid architecture against a ``structural twin'' classical baseline. These baselines replace the quantum components with standard classical layers of equivalent dimensionality, while keeping the rest of the architecture identical.

\begin{itemize}
    \item In our baseline for HQNN-Parallel, the quantum linear layer branch is bypassed. The feature vector is processed solely by a classical Multi-Layer Perceptron (MLP) consisting of three fully connected layers with decreasing hidden sizes. LeakyReLU activations are applied after the first two layers to introduce non-linearity comparable to the quantum variance.
    
    \item In our baseline for HQNN-Quanv, the initial quantum convolution layer is replaced by a standard classical \texttt{Conv2d} layer. To ensure fair comparison regarding spatial dimension reduction, we employed a kernel size of $3\times3$, a stride of $2$, and padding of $1$. This configuration matches the effective receptive field and output dimensionality of the quantum kernel approach.
\end{itemize}

It is important to note that these baselines are controlled architectural references, not attempts to achieve State-of-the-Art (SOTA) results on EuroSAT, which typically requires large-scale pre-trained models.

To ensure the comparison relies solely on architecture rather than optimization tuning, both quantum and classical variants were trained using a centralized trainer with identical hyperparameters. We utilized the \texttt{AdamWScheduleFree} optimizer \cite{defazioRoadLessScheduled2024} with a learning rate of $0.01$ and trained for $10$ epochs. Batch sizes were consistent within each experiment type ($16$ for HQNN-Parallel and $4$ for HQNN-Quanv).

\subsection{Experiments}\label{experiments}
For each architecture, we systematically evaluate all possible combinations of encoding, ansatz, and measurement methods under identical learning rates, batch sizes, and dataset parameters. These are benchmarked against the classical baselines defined in the previous section.

This approach aims to isolate and assess the impact of architectural design choices while maintaining consistent hyperparameters. Also, it aims to ensure that performance differences are attributable to the quantum architectural components rather than training variance. Although further optimization could enhance results, the primary objective of this study is to analyze the effects of these design decisions rather than pursuing performance maximization. See more details in \hyperref[table1]{TABLE \ref*{table1}} for the different encodings, ans\"atze and measurements used for each model.

The dataset comprises $500$ samples with an $80/20$ split, leading to $50$ examples per class. The hybrid architectures use both $32\times32$ and $16\times16$ pixel resolutions. The QCNN model exclusively utilizes $16\times16$ images to streamline computational requirements while maintaining essential feature extraction capabilities.

\begin{table*}[htbp!]
\begin{tabular}{lllllllll}
\hline
\textbf{Model} &  & \textbf{Encodings} &  & \textbf{Variational ans\"atze} &  & \textbf{Measurements} &  & \textbf{Classical processing}                  \\ \hline
\textit{}      &  &                    &  &                              &  &                       &  &                                                \\
\textit{}      &  & - Angle            &  & - No entanglement               &  & - Pauli X             &  & The model initially employs                    \\
               &  & - Amplitude        &  & - Fully entangled             &  & - Pauli Y             &  & classical convolutional layers                 \\
\textit{HQNN}           &  & - IQP              &  & - Ring                       &  & - Pauli Z             &  & for feature extraction, followed               \\
\textit{Parallel}       &  & - QAOA             &  & - NQ                         &  &                       &  & by a quantum circuit integrated                \\
               &  & - Ring             &  &                              &  &                       &  & within the fully connected stage               \\
               &  & - Waterfall        &  &                              &  &                       &  &                                                \\
               &  &                    &  &                              &  &                       &  &                                                \\
               &  &                    &  &                              &  &                       &  &                                                \\
\textit{}      &  & - Angle            &  & - No entanglement               &  & - Pauli X             &  & After the \emph{quanvolution} \\
               &  & - Amplitude        &  & - Fully entangled             &  & - Pauli Y             &  & layer, the model proceeds with                 \\
\textit{HQNN}           &  & - IQP              &  & - Ring                       &  & - Pauli Z             &  & a classical convolutional layer                \\
\textit{Quanv}         &  & - QAOA             &  & - NQ                         &  &                       &  & to further extract features,                   \\
               &  & - Ring             &  &                              &  &                       &  & followed by fully connected                    \\
               &  & - Waterfall        &  &                              &  &                       &  & layers for classification                      \\
               &  &                    &  &                              &  &                       &  &                                                \\
\textit{}      &  &                    &  &                    &  & - Pauli Z               &  &                                                \\
\textit{QCNN}  &  & - Amplitude          &  &     - QCNN               &  & - Paulis            &  & None                                           \\
               &  &                    &  &                     &  & - Histogram             &  &                                                \\
               &  &                    &  &                              &  &                       &  &                                                \\
               &  &                    &  & - Two kernel                   &  & - Pauli Z               &  &                                                \\
\textit{SEQNN} &  & - Amplitude          &  & - FC (Simplified Two Design)                   &  & - Paulis            &  & None                                           \\
\textit{}      &  &                    &  &                     &  & - Histogram             &  &                                                \\
\textit{}      &  &                    &  &                              &  &                       &  &                                                \\ \hline
\end{tabular}
\caption{Summary of the models and their main components.}
\label{table1}
\end{table*}

\subsection{Metrics and optimization}

We train our models using the schedule-free variant of the AdamW optimizer \cite{defazioRoadLessScheduled2024} coupled with the standard Cross Entropy Loss \cite{CEL}. For evaluation, we report classification accuracy throughout the experiments. Additional metrics (e.g., precision, recall, F1-score) were computed using the scikit-learn library \cite{sklearn}, and detailed definitions can be found in its documentation.

In all training runs we used the following hyperparameters: $0.01$ learning rate, $16$ batch size and $30$ epochs with $10$ epoch patience. 

\section{Results}\label{sec:results}

The data collected during these experiments provided critical insights into the system's performance under varying conditions, thereby supporting the subsequent discussion of the results.

\subsection{HQNN-Parallel}

By selecting the best validation accuracy results for each encoding-ans\"atze combination, we observe that several hybrid configurations match or slightly exceed the classical baseline, as shown in \hyperref[parallel-best-acc]{FIG. \ref*{parallel-best-acc}}. For instance, with \(32 \times 32\) images, the IQP encoding combined with the no entanglement variational circuit achieves \(75.1\%\) accuracy, outperforming the classical model's \(67.1\%\).

In \hyperref[parallel-learning]{FIG. \ref*{parallel-learning}}, we depict the learning behavior of the best-performing hybrid network and compare it with the classical model. The validation accuracy (\hyperref[parallel-learning]{FIG. \ref*{parallel-learning}c}) shows that the best hybrid model achieves better results than the classical one. However, the validation loss (\hyperref[parallel-learning]{FIG. \ref*{parallel-learning}d}) indicates that both models begin to overfit after a few epochs, as the loss starts to increase. Prolonged training would likely exacerbate this issue. 

We will now analyze how each component of the circuit influences the final performance.

\subsubsection*{How does the encoding influence?}\label{hqnnencoding}

\begin{figure}[!htbp]
 \includegraphics[width=1\columnwidth]{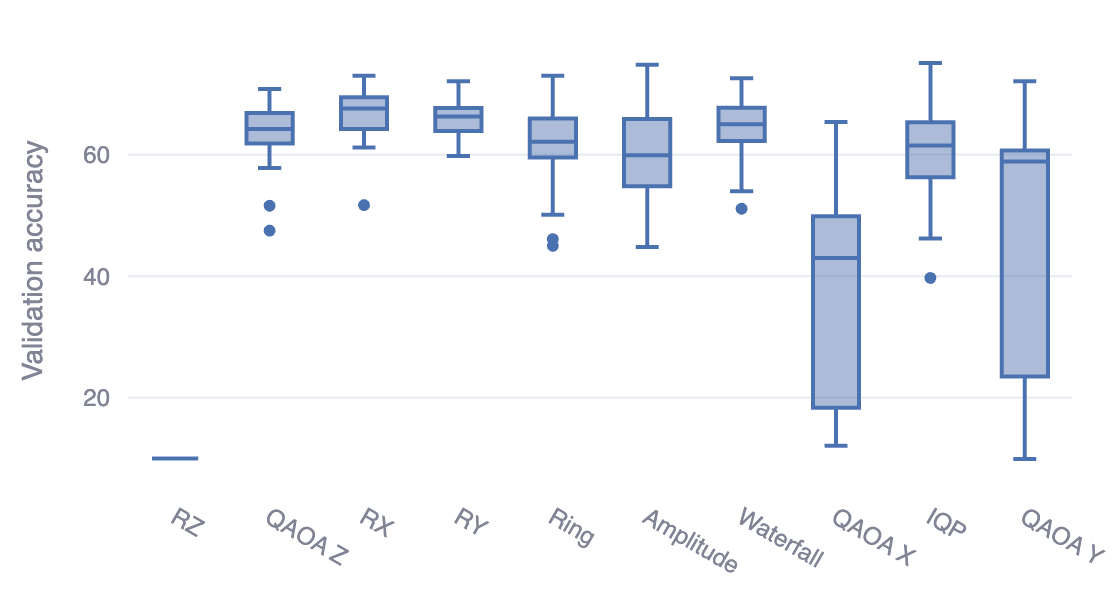}
 \caption{Validation accuracy of all configurations for a fixed encoding.}
 \label{parallel-encoding-effect}
\end{figure}

Regarding the encoding method, no single solution is universally optimal. However, as seen in \hyperref[parallel-encoding-effect]{FIG. \ref*{parallel-encoding-effect}}, the $RX$, $RY$ and $QAOA$ $Z$ encodings exhibit lower variance in their performance. In contrast, the $Z$ rotation-based encoding ($RZ$) consistently fails, yielding only \(10\%\) accuracy. As mentioned in the description of angle encoding, this is due to the encoding only adding an irrelevant global phase. 

Interestingly, while $QAOA$ $Z$ is among the best performing strategies, the encodings $QAOA$ $X$ and $QAOA$ $Y$ can fail unpredictably. Numerical analyses (see \hyperref[appendix:expressibility]{Appendix \ref*{appendix:expressibility}} ) suggest that $QAOA$ $X$ and $QAOA$ $Y$ have higher expressibility than their analogues with a local field in the $Z$ direction. It has been already established in the field that an excess of expressibility can hamper the trainability of a circuit \cite{Larocca2025}. Hence, depending on where the optimization is initialized, the optimizer may or may not be able to properly adjust the circuit parameters in order to reach a high accuracy, yielding a big variance in performance, as seen in \hyperref[parallel-encoding-effect]{FIG. \ref*{parallel-encoding-effect}}.

The box plots illustrate the distribution of validation accuracy across the different configurations tested for each category. The central horizontal line indicates the median, the box represents the interquartile range between the 25th and 75th percentiles, and the whiskers extend to the minimum and maximum values excluding outliers, which are plotted as individual dots. Each data point within a category corresponds to a unique combination of encoding and ansatz. While each specific configuration was executed once, preliminary sensitivity analyses confirmed low variance across different parameter initializations, indicating that the reported results are representative of the model's stable performance.

\subsubsection*{How does the ansatz influence?}

For the variational part of the circuit, we find that all tested ans\"atze yield similar performance, as illustrated in \hyperref[parallel-ansats-effect]{FIG. \ref*{parallel-ansats-effect}}. The median and maximum accuracies are comparable across all configurations. This suggests that for this particular problem, the choice of entangling structure within the ansatz is not a critical factor for achieving high accuracy.

\subsubsection*{How does the measurement influence?}

The choice of measurement basis also shows a relatively small impact on the overall performance. As shown in \hyperref[parallel-meas-effect]{FIG. \ref*{parallel-meas-effect}}, all measurement strategies produce similar accuracy distributions. There is a slight tendency for measurements in the computational basis to perform better than measurements in the X or Y bases, but the difference is not substantial.

\subsection{HQNN-Quanv}

By selecting the best validation accuracy results for each encoding-ansatz combination, we observe that none of the outcomes surpass the classical case. However, as illustrated in \hyperref[allflex]{FIG. \ref*{allflex}}, certain combinations exhibit performance very close to the classical baseline. Although these results are derived in a simulated context, they provide insights into the types of circuits that should be considered in QML. We will analyze how each circuit type influences the outcomes.

In \hyperref[fig:flex_metrics_val]{FIG. \ref*{fig:flex_metrics_val}} depicts the best network's learning behavior and corresponding losses compared with the classical ones. As shown in the validation loss, there is a point at which both the classical and hybrid models cease to learn, leading to an increase in loss. Furthermore, our analysis indicates that the train performance continues to improve during training. However, allowing additional training epochs may lead to overfitting, as prolonged learning could cause the model to excessively adapt to the training data rather than generalizing effectively.

\subsubsection*{How does the encoding influence?}
\begin{figure}[!htbp]

 \includegraphics[width=1\columnwidth]{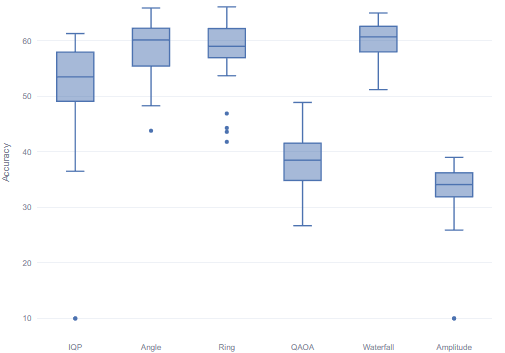}
 \caption{Validation accuracy comparison across encodings. Image shapes ($32\times32$ and $16\times16$) and quantum kernel shapes (qks = 2, 3) are included.}
    \label{flexencoding}
\end{figure}

As we can see in \hyperref[flexencoding]{FIG. \ref*{flexencoding}}, the amplitude and QAOA encodings fail to capture relevant features. In contrast, Waterfall, Ring and Angle encodings demonstrate greater utility for this task, although they yield slightly inferior performance compared to the classical approach. These findings suggest that, when quantum information is available, the aforementioned encodings should be prioritized for implementing \emph{quanvolutions} of this type. The IQP encoding occupies an intermediate position, indicating potential scenarios where its application could be advantageous.

In this case we excluded $Z$ rotations of angle encoding because they only add an irrelevant global phase, as mentioned in section \hyperref[hqnnencoding]{II.A.1} .

\subsubsection*{How does the ansatz influence?}
Non entangled circuits are less effective than entangled ones. In \hyperref[flex_ansatz]{FIG. \ref*{flex_ansatz}}, the bulk of accuracies for the non-entangled ones obtains a worse performance than for the other 3 that do have at least some entanglement. Through this approach, we observe that image features are more effectively captured by leveraging the phenomenon of entanglement.

\subsubsection*{How does the measurement influence?}

Preliminary results suggest that the type of measurement employed does not significantly affect the outcomes, as similar values are observed when evaluating the results. See \hyperref[flex_measurement]{FIG. \ref*{flex_measurement}}.

\subsubsection*{How does the quantum kernel shape influence?}

Before conducting the experiments, we expected substantial differences in results depending on the qks, as the circuits employed qks$^{2}$ qubits. However, it was not the case; we see similar accuracies for qks $=2,3$, maybe a little better for shape 3 (\hyperref[flex_qks]{FIG. \ref*{flex_qks}}). The results for a qks of size 5 remain pending due to the high computational cost associated with a 25-qubit simulated circuit.

\subsubsection*{How does the image size influence?}
Results suggest that image size does not significantly impact the model's accuracy. This observation highlights a potential advantage, as reduced-resolution images appear to retain sufficient discriminative information comparable to higher-resolution counterparts. However, it is critical to acknowledge that these experiments utilized extremely small images, which may limit the broader applicability of this conclusion and warrant further investigation under diverse image-scale conditions. See \hyperref[flex_imgsize]{FIG. \ref*{flex_imgsize}}.

\subsection{QCNN}

\subsubsection*{How does the encoding influence?}
Changing the ordering in the amplitude embedding to random tends to decrease the accuracy (See \hyperref[appendix_qcnn_orderings]{FIG. \ref*{appendix_qcnn_orderings}}). This is less evident in the QCNN and FC architectures, as they were not designed to exploit this property. We are only showcasing here the best performing ordering against a random one, which we considered a baseline.  

Although our \emph{FC} model is supposed to mimic an unstructured, fully connected layer, it is composed of adjacent two-qubit gates; hence, it is also affected by the change in ordering. 

\subsubsection*{How does the measurement influence?}
Within each group of measurements used for the quantum architectures, there is still room for some flexibility. In the \emph{PauliZ} and \emph{histogram} types of measurements, in some cases there are more active qubits than needed at the end; and in the \emph{Paulis}, when measuring $N$ qubits, we have in general $4^N-1$ non-trivial Pauli operators to select from. The choice of operator can greatly influence the classification accuracy as well; however, we are picking one at random in the study due to the vast range of possibilities.

In general, the best performing type of measurement was the \emph{PauliZ}, which required measuring as many qubits as classes. Pauli measurement gives the most flexibility at the cost of a great dispersion depending on the chosen operators (\hyperref[appendix_qcnn_measure]{FIG. \ref*{appendix_qcnn_measure}}).

\subsubsection*{How does the FC layer depth influence?}
We also tested whether increasing the number of parameters of the circuits by making the \emph{FC} layers deeper influenced the performance of the models. Interestingly, we saw no improvement when increasing the number of parameters for 4 classes, suggesting that architectural changes are needed to enhance model classification capabilities (see \hyperref[appendix_qcnn_fc_layer_depth]{FIG. \ref*{appendix_qcnn_fc_layer_depth}}). However, for 10 classes, increasing the depth of the \emph{FC} layer did have a positive impact on the accuracy. A brief look at the training curves (\hyperref[qcnn_metrics]{FIG. \ref*{qcnn_metrics}}) points towards the fact that we are indeed well in the underparameterization regime (all purely quantum models have less than 400 parameters).

\section{Evaluation}\label{sec:evaluation}

This section reflects on the qualitative lessons that emerged once the quantitative sweep of circuit configurations was complete, explaining \emph{why} certain components matter and how they interact. 

\subsection{Hyperparameters}\label{laxd}
In this work, we did not consider adjusting the optimization hyperparameters due to the associated computational cost. 
Optimizer hyperparameters such as the learning rate, momentum, and adaptivity coefficients can critically affect the geometry of the loss landscape. A large learning rate in high curvature regions may cause oscillations or even divergence, while flat regions can lead to slow (or lack of) convergence if explored with a small learning rate. Similarly, other hyperparameters such as batch size and gradient averaging or momentum control the stochasticity in parameter updates. A loss landscape with many local minima may benefit from a higher effective noise level parameter update, while a smoother loss profile is more efficiently optimized under a more stable parameter update strategy. Additionally, it would be beneficial to investigate how these hyperparameters affect metrics beyond validation accuracy, such as generalization and trainability. Optimizer performance and convergence behavior are inherently landscape-dependent, making systematic hyperparameter sensitivity analysis a nontrivial and computationally demanding task that was out of the scope of the present work.

Regarding trainability, while an empirical analysis of gradient variance was not performed due to current constraints in scaling circuits to a higher number of qubits, existing literature provides a strong theoretical foundation for our approach. Specifically, fundamental studies \cite{PhysRevX.11.041011} have analytically demonstrated that QCNN-type architectures are exempt from the Barren Plateau (BP) phenomenon, exhibiting only polynomial gradient vanishing, as long as they maintain logarithmic depth and employ local measurements. For our remaining architectures, we mitigate BP risks by maintaining shallow circuit depths and utilizing local observables, which are recognized strategies to preserve gradient signal in high-dimensional spaces \cite{Larocca2025}.

\subsection{Applicability to Other Datasets} We employed the EuroSAT dataset \cite{Helber2019} as a rigorous testbed for our proposed models. EuroSAT was specifically chosen because it offers a necessary balance for NISQ-era quantum simulation: it presents significant classification challenges (high intra-class variance and complex textures) while maintaining a resolution that allows for manageable downsampling without significant feature loss. This makes it superior to simpler datasets, like MNIST variants, for evaluating model performance. Although we anticipate that our results would extrapolate to other types of imagery with similar behaviors, the high computational cost of simulating quantum circuits limits the feasibility of multi-dataset benchmarking in this study. Consequently, the inclusion of further remote sensing datasets remains a priority for future work.

\subsection{Position of the quantum layer and effective entanglement}

Hybrid models present two distinct forms of adding the quantum layers to classical networks. 

When the layer is positioned in the latter part of the model, it operates on the features extracted by the earlier layers. With this approach, we found that the performance was slightly better than the classical counterpart, suggesting that increasing the qubits allocated per circuit, which would enable long-range correlations between features, could allow for better potential performance.

Conversely, situating the PQC-based convolutional layer at the beginning of the model led to a marginally worse performance than its classical counterpart. Suggesting that operating directly on raw pixels might not be optimal and alternative representations or preprocessing should be explored.

We found entanglement to be necessary for achieving top validation accuracy performance for the HQNN-Quanv architecture. In all cases where a non-entangling encoding and ansatz were used, the performance was worse than the entangled alternatives. For the HQNN-Parallel, in some cases the non-entangled combination of ansatz and encoding was slightly better than some entangled versions. However, we would expect entangling combinations to perform better for higher qubit counts, as in our experiments each circuit only had access to $8/768$ features.

These observations should be taken with a grain of salt, given that for the hybrid models we only measured local observables, and outcomes could be very different for other types of measurements.

\subsection{Encoding, measurement, and their interplay}

For both hybrid models, we found the encoding to be the most determining factor,  \hyperref[parallel-encoding-effect]{FIG. \ref*{parallel-encoding-effect}}, \hyperref[flexencoding]{\ref*{flexencoding}}. Exceeding the effect of both ansatz and measurement, \hyperref[parallel-encoding-effect]{FIG. \ref*{parallel-measurement-effect}}, \hyperref[flexencoding]{\ref*{appendix_ans_measure}}. 

We speculate this is due to several factors. First, the encoding is known to be a double-edged sword: low-complexity encodings don't have enough expressive power \cite{schuldEffectDataEncoding2021} while excessively complex encodings lead to data concentration \cite{liConcentrationDataEncoding}, which limits the distinguishability of encoded states. Secondly, we tried few ans\"atze compared to encodings. Perhaps, by adding a range of carefully selected options with varying complexities, the impact would be more clear. Lastly, local measurements on Pauli $X/Y/Z$ may be too simple to meaningfully affect performance.

Amplitude encoding thrived only when the circuit processed a compressed global representation; it faltered whenever it was fed directly with small pixel windows, because normalizing entire patches dilutes the very gradients that early layers should preserve. 

Single-qubit rotation encodings, by contrast, maintained robust performance regardless of placement, showing that locality in the data-to-state map is valuable whenever the network itself is still local.

For hybrids, the measurements were significantly less decisive, which is logical as they are highly similar to each other and can be interconverted via single-qubit rotations. Therefore, three Cartesian single-qubit bases produced practically identical scores, implying that the subsequent classical layers rotate whatever Bloch-vector projection they are given.  In a purely quantum network, where no further classical post-processing is available until the final softmax, the simplest global measurement in the computational basis offered the clearest separation of classes, whereas multiqubit Pauli strings introduced unnecessary complexity without a matching gain. The study was not particularly exhaustive in this aspect, and given that these measurements are more distinct from one another than in the previous case, it is reasonable to expect increased variability among them.

Interestingly, the fully quantum architectures used only amplitude encoding, yet a simple reordering of amplitudes before the first convolution layer delivered a noticeable lift.  By arranging neighboring pixels so that two-qubit gates act on semantically related amplitudes, the circuit effectively implements shaped kernels, mirroring how classical convolutions exploit spatial proximity.

\subsection{Practical implications and open questions}

When validation accuracy is plotted against the number of trainable parameters, the classical baselines reveal the familiar pattern of saturating returns.  The fully quantum models, however, achieve the same operating point while maintaining an order-of-magnitude smaller parameter budget.  Although present-day simulation makes wall-clock training slower, the marked reduction in learnable weights hints at a tangible memory advantage for future quantum accelerators. The promising aspect here is the remarkably small size of the model and the significant results it can achieve with so few parameters. The accuracy-parameter ratio is substantially higher in the QCNN compared to all others as we can see in \hyperref[table2]{TABLE \ref{table2}}.

\begin{table}[h]
\centering
\begin{tabular}{llrlrrr}
\hline
\textbf{Model} &  & \multicolumn{1}{l}{\textbf{\makecell[l]{No. of \\ params}}} &  & \multicolumn{1}{l}{\textbf{Accuracy (\%)}} & \multicolumn{1}{l}{} & \multicolumn{1}{l}{\textbf{\makecell[l]{Accuracy- \\ Param $\cdot$ 100}}} \\ \hline
\textit{}              &  & \multicolumn{1}{l}{}                       &  & \multicolumn{1}{l}{}                  & \multicolumn{1}{l}{} & \multicolumn{1}{l}{}                         \\
\textit{\makecell[l]{Classical \\ "Parallel"}} &  & $903598$                                 &  & $67.1$                                &                      & $0.007$                                      \\
                       &  &                                            &  &                                       &                      &                                              \\
\textit{\makecell[l]{Classical \\ "Quanv"}}    &  & $273706$                                 &  & $71.5$                                &                      & $0.026$                                      \\
                       &  &                                            &  &                                       &                      &                                              \\
\textit{HQNN-Parallel} &  & $15658$                                      &  & $74.8$                                     &                      & $0.477$                                       \\
                       &  &                                            &  &                                       &                      &                                              \\
\textit{HQNN-Quanv}    &  & 930826                                     &  & $65.9$                                  &                      & $0.007$                                      \\
\textit{}              &  &                                            &  &                                       &                      &                                              \\
\textit{QCNN}          &  & $300$                                      &  & $42.5$                                &                      & $15$                                         \\
\textit{}              &  & \multicolumn{1}{l}{}                       &  & \multicolumn{1}{l}{}                  & \multicolumn{1}{l}{} & \multicolumn{1}{l}{}                         \\ \hline
\end{tabular}
\caption{Comparison of parameter count and accuracy-parameter ratio across models. For all hybrid-quantum models, the optimal configuration was selected. HQNN-Parallel: Amplitude-No Entanglement; HQNN-Quanv: AngleX-Full Entanglement; QCNN-Flatten-Paulis.}
\label{table2}
\end{table}

Several avenues remain open.  First, the sweet-spot in entanglement may drift once realistic hardware noise is included, so repeating the sweep on early fault-tolerant devices would clarify the hardware-aware optimum. Moreover, analyzing the non-stabilizerness of the states generated by the different encodings and ansatze may shed light on which combinations are best suited to provide value regarding the classical counterpart. Efforts are already being made to study the behavior of such metrics in the context of PQCs \cite{capecci2025role}. Second, data re-uploading layers—whose parameters modulate the encoding itself—could compensate for the shortcomings of an early quanvolution by learning a task-specific Fourier spectrum.  Finally, a systematic survey of measurement strategies under realistic noise models is still lacking; understanding which observables remain informative when decoherence is present will be key to deploying quantum classifiers in practice.

It is crucial to interpret these results within the context of ideal, noiseless simulations. While our analysis identifies a performance hierarchy where entangled ans\"atze generally outperform non-entangled variants (particularly in the HQNN-Quanv architecture), this advantage may be compromised in a real-world NISQ environment. On physical hardware, the introduction of entanglement increases circuit depth and gate count, thereby accumulating decoherence and operational errors. Consequently, a trade-off emerges where the theoretical expressibility gained by entanglement must be weighed against the hardware-induced noise. Future work must rigorously validate these findings under realistic noise models to determine the threshold at which the benefits of entanglement are negated by gate errors.

\section{Conclusion}

The study addresses the challenge of understanding how design choices in PQCs, such as data encoding, variational ans\"atze, and measurement bases, affect the performance of quantum and hybrid neural network architectures in image classification tasks using the EuroSAT dataset \cite{Helber2019}.

Hybrid models are most sensitive to data encoding strategies, with amplitude-based embeddings being incompatible with certain architectures (e.g., HQNN-Quanv) but effective in others (e.g., HQNN-Parallel). Variational ans\"atze and measurement bases have a secondary impact, but the entangled ans\"atze performed better in HQNN-Quanv than HQNN-Parallel. Purely quantum models, constrained to amplitude encoding, show performance dependence on the combination of measurement strategies and input ordering. The latter models are considerably smaller than the current classical models.
  
Beyond merely enhancing current results, purely quantum models hold the promise of achieving exceptionally high accuracy-parameter ratios compared to existing methods. With adequate resources, this could lead to significantly lighter models with reduced computational overhead. Such advancements would enable more efficient and scalable QML applications, making them accessible for a broader range of practical use cases. This shift could revolutionize fields such as image recognition, natural language processing, and complex data analysis.
  
We are limited to a specific dataset and predefined PQC configurations, which may not generalize to other tasks or quantum hardware. Additionally, the evaluation does not account for potential hardware-specific constraints or noise effects in quantum computations.
  
As future work, it would be interesting to consider the aforementioned points, like adjusting hyperparameters or even exploring different image datasets. On the other hand, a thorough analysis of the parameter-accuracy relationship will be crucial to select a suitable model for implementation in quantum computers.
  
This study not only identifies determinants in the performance of hybrid and quantum models, but also highlights how well-designed quantum circuits can overcome traditional barriers, bringing us closer to a future where QML is accessible and effective in complex industrial problems.

\section*{Acknowledgments}

The authors would like to express their gratitude for the collaborative efforts within the ARQA Cervera Technology Transfer Network \cite{ARQA}. This work is a result of the collective contributions and support from all members of this network, which has significantly enhanced our research outcomes.

OB is a fellow of Eurecat’s ``Vicente L\'opez'' PhD grant program.

We would like to thank Andr\'es Garc\'ia Mangas and Marco Aguado Acevedo for his support in launching and executing the experiments using the \textit{CTIC Quantum Testbed (QUTE)} platform \cite{qute}.

\section*{Code and data availability}

All code and data used to generate the results presented in this paper are publicly available on GitHub at \url{https://github.com/uriballo/QML-Satellite-Image-Classification}.




\onecolumngrid
\vspace{0.75cm}
\hrule

\bibliography{paper}

\newpage
\appendix
\makeatletter
\def\@oddhead{APPENDIX \hfill \thepage}
\def\@evenhead{APPENDIX \hfill \thepage}
\makeatother

\newpage
\section{Shaped Embedding Quantum Neural Network, detailed explanation}\label{appendix:SEQNN}
In this section we will show how we can use the action of a two-qubit gate acting on adjacent qubits to implement convolutional kernels over the original images by selecting the way in which pixels are introduced in the amplitudes of the state vector. 

\subsection{The math behind}
The action of a general two qubit gate is given by
\begin{equation}
    \mathcal{G} \, |\varphi\rangle =
\begin{pmatrix}
g_{00} & g_{01} & g_{02} & g_{03} \\
g_{10} & g_{11} & g_{12} & g_{13} \\
g_{20} & g_{21} & g_{22} & g_{23} \\
g_{30} & g_{31} & g_{32} & g_{33}
\end{pmatrix}
\begin{pmatrix}
\varphi_0 \\
\varphi_1 \\
\varphi_2 \\
\varphi_3
\end{pmatrix}
=
\begin{pmatrix}
\vec{g}_0 \cdot \vec{\varphi} \\
\vec{g}_1 \cdot \vec{\varphi} \\
\vec{g}_2 \cdot \vec{\varphi} \\
\vec{g}_3 \cdot \vec{\varphi}
\end{pmatrix}
\end{equation}

where $\vec{\varphi} = (\varphi_0, \varphi_1, \varphi_2,\varphi_3) $ and $\vec{g}_i$ corresponds to the i-th row of $\mathcal{G}$. Thus, each row of the matrix representing the gate can be seen as a kernel acting on the four entries of the two-qubit state vector. Now, if we have three qubits and perform a gate on the two right-most ones,

\begin{equation}
\mathbb{I}_0 \otimes \mathcal{G}_{1,2} |\varphi\rangle=
\begin{pmatrix}
\mathcal{G} & \mathbf{0}_{2 \times 2} \\
\mathbf{0}_{2 \times 2} & \mathcal{G}
\end{pmatrix}
|\varphi\rangle =
\begin{pmatrix}

\begin{pmatrix}
\vec{g}_0 \\
\vec{g}_1 \\
\vec{g}_2 \\
\vec{g}_3
\end{pmatrix}
\cdot \vec{\varphi}_0

& \mathbf{0}_{2 \times 2} \\
\mathbf{0}_{2 \times 2} &
\begin{pmatrix}
\vec{g}_0 \\
\vec{g}_1 \\
\vec{g}_2 \\
\vec{g}_3
\end{pmatrix}
\cdot \vec{\varphi}_1
\end{pmatrix}
\end{equation}
where $\vec{\varphi}_0 = (\varphi_0, \varphi_1, \varphi_2,\varphi_3) $, $\vec{\varphi}_1 = (\varphi_4, \varphi_5, \varphi_6,\varphi_7) $. The full 8-element state vector is arranged in groups of four pixels which are mixed together by the corresponding kernels associated with the rows of the gate. 

If the gate acts on the left-most qubits, we are still implementing kernels that mix amplitudes of the state vector, but now these groups are not comprised of subsequent amplitudes. More explicitly: 
\begin{equation}
    \mathcal{G}_{0,1} \otimes \mathbb{I}_2 =
\begin{pmatrix}
g_{00} \begin{pmatrix} 1 & 0 \\ 0 & 1 \end{pmatrix} & g_{01} \begin{pmatrix} 1 & 0 \\ 0 & 1 \end{pmatrix} & g_{02} \begin{pmatrix} 1 & 0 \\ 0 & 1 \end{pmatrix} & g_{03} \begin{pmatrix} 1 & 0 \\ 0 & 1 \end{pmatrix} \\
g_{10} \begin{pmatrix} 1 & 0 \\ 0 & 1 \end{pmatrix} & g_{11} \begin{pmatrix} 1 & 0 \\ 0 & 1 \end{pmatrix} & g_{12} \begin{pmatrix} 1 & 0 \\ 0 & 1 \end{pmatrix} & g_{13} \begin{pmatrix} 1 & 0 \\ 0 & 1 \end{pmatrix} \\
g_{20} \begin{pmatrix} 1 & 0 \\ 0 & 1 \end{pmatrix} & g_{21} \begin{pmatrix} 1 & 0 \\ 0 & 1 \end{pmatrix} & g_{22} \begin{pmatrix} 1 & 0 \\ 0 & 1 \end{pmatrix} & g_{23} \begin{pmatrix} 1 & 0 \\ 0 & 1 \end{pmatrix} \\
g_{30} \begin{pmatrix} 1 & 0 \\ 0 & 1 \end{pmatrix} & g_{31} \begin{pmatrix} 1 & 0 \\ 0 & 1 \end{pmatrix} & g_{32} \begin{pmatrix} 1 & 0 \\ 0 & 1 \end{pmatrix} & g_{33} \begin{pmatrix} 1 & 0 \\ 0 & 1 \end{pmatrix}
\end{pmatrix}
=
\begin{pmatrix}
g_{00} & 0 & g_{01} & 0 & g_{02} & 0 & g_{03} & 0 \\
0 & g_{00} & 0 & g_{01} & 0 & g_{02} & 0 & g_{03} \\
g_{10} & 0 & g_{11} & 0 & g_{12} & 0 & g_{13} & 0 \\
0 & g_{10} & 0 & g_{11} & 0 & g_{12} & 0 & g_{13} \\
g_{20} & 0 & g_{21} & 0 & g_{22} & 0 & g_{23} & 0 \\
0 & g_{20} & 0 & g_{21} & 0 & g_{22} & 0 & g_{23} \\
g_{30} & 0 & g_{31} & 0 & g_{32} & 0 & g_{33} & 0 \\
0 & g_{30} & 0 & g_{31} & 0 & g_{32} & 0 & g_{33}
\end{pmatrix}
\end{equation}

\begin{equation}
    \mathcal{G}_{0,1} \otimes \mathbb{I}_2 |\varphi\rangle =
\mathcal{G}_{0,1} \otimes \mathbb{I}_2
\begin{pmatrix}
\varphi_0 \\
\varphi_1 \\
\varphi_2 \\
\varphi_3 \\
\varphi_4 \\
\varphi_5 \\
\varphi_6 \\
\varphi_7
\end{pmatrix}
=
\begin{pmatrix}
\vec{g}_0 \cdot \vec{\varphi}_0 \\
\vec{g}_0 \cdot \vec{\varphi}_1 \\
\vec{g}_1 \cdot \vec{\varphi}_0 \\
\vec{g}_1 \cdot \vec{\varphi}_1 \\
\vec{g}_2 \cdot \vec{\varphi}_0 \\
\vec{g}_2 \cdot \vec{\varphi}_1 \\
\vec{g}_3 \cdot \vec{\varphi}_0 \\
\vec{g}_3 \cdot \vec{\varphi}_1
\end{pmatrix}
\end{equation}

and $\vec{\varphi}_0 = (\varphi_0, \varphi_2, \varphi_4,\varphi_6) $, $\vec{\varphi}_1 = (\varphi_1, \varphi_3, \varphi_5,\varphi_7) $.

\textbf{Conclusion:} for $N+1$ qubits, there are $2^{N+1}$ amplitudes, and $2^{N-1}$ groups of 4 amplitudes that are being mixed together by the action of a gate operating on two adjacent qubits. If the two-qubit gate is in position $p$, the $2^{N-1}$ groups of 4 amplitudes are:
\begin{equation}
    \vec{\varphi}_l^{(p)} = (\varphi_{l},\varphi_{l+2^p},\varphi_{l+2\cdot 2^p},\varphi_{l+3\cdot 2^p}), \quad l \in \{0,\ldots,2^{N-1}\}.
\end{equation}

\begin{figure}[h!]
    \centering
    \includegraphics[width=0.5\linewidth]{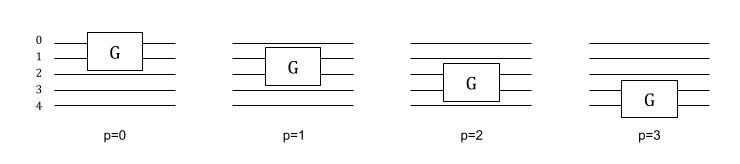}
    \caption{The four possible positionings of a two-qubit gate acting on adjacent qubits of a 5-qubit register.}
    \label{fig:two-qubit-gate}
\end{figure}

It is worth noting that with the same kind of reasoning we can see that a general unitary acting of three adjacent qubits can implement a kernel that mixes $2^3=8$ amplitudes of the state vector and, in general, we can create a kernel over $2^K$ pixels with a gate operating on $K$ adjacent qubits.

\subsection{Custom orderings}
As a first attempt to try to test the idea explained in the last section, we designed specific orderings for the initial amplitude embedding such that it would leverage the action of adjacent two qubit gates to implement some kernels. 

\begin{figure}[h!]
    \centering
    \begin{subfigure}[t]{0.47\textwidth}
        \centering
        \includegraphics[width=\textwidth]{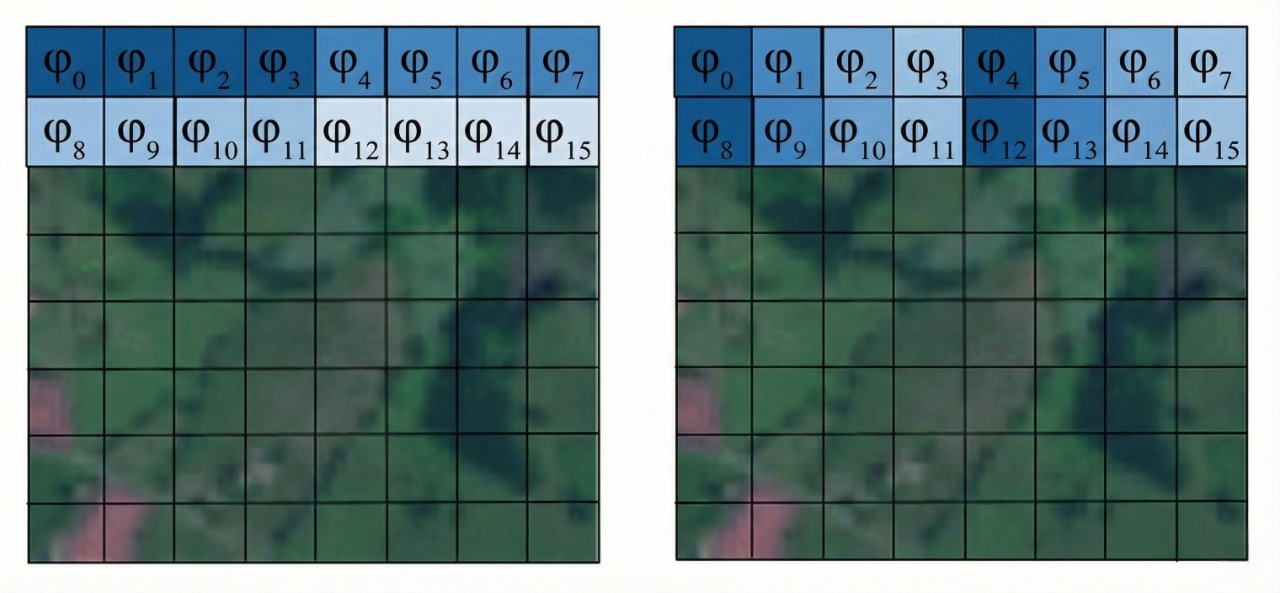}

        \caption{Flattening the image array in a row first fashion corresponds with a kernel that acts on horizontal lines (first gate) and mixes every fourth pixel (second gate).}
        \label{fig:flatten}
    \end{subfigure}
    \hspace{0.04\textwidth}
    \begin{subfigure}[t]{0.47\textwidth}
        \centering
        \includegraphics[width=\textwidth]{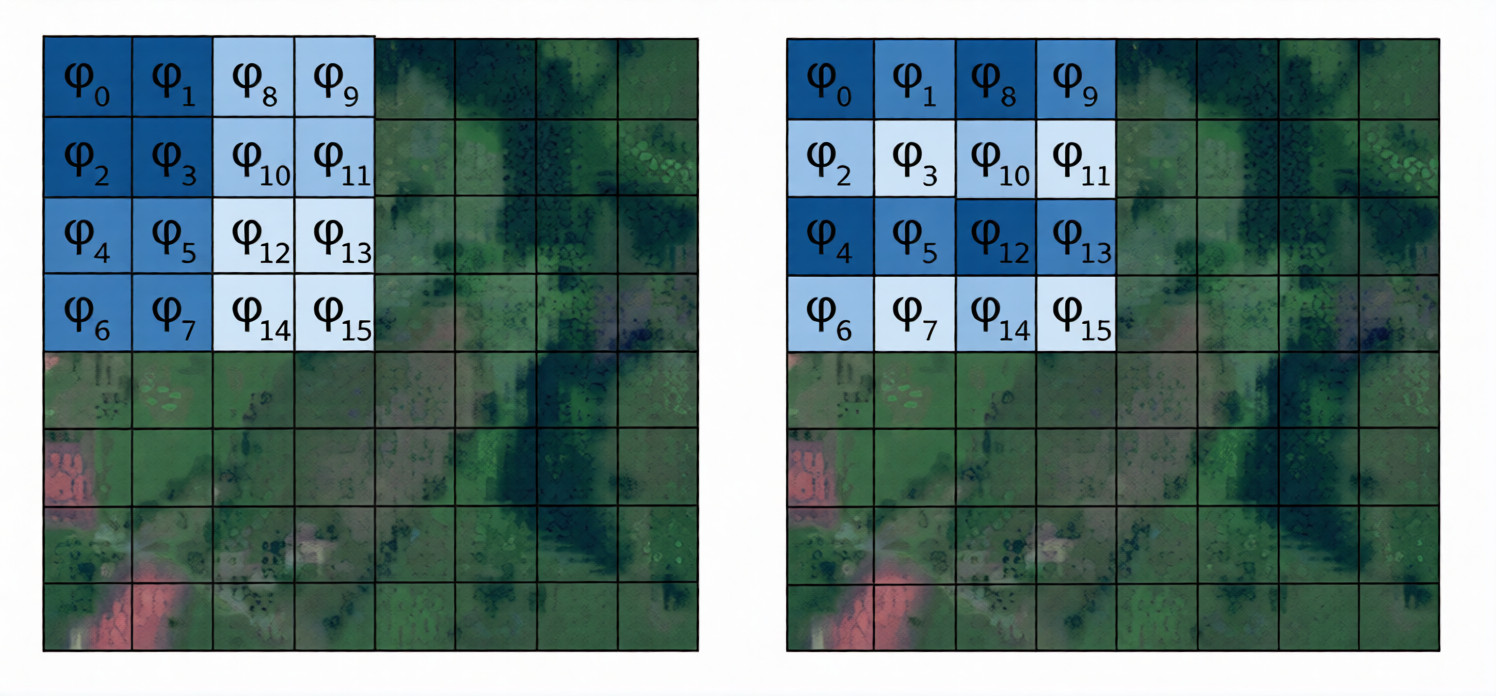}
        \caption{This ordering implements a kernel that acts on squared regions of the images.}
        \label{fig:2x2blocks}
    \end{subfigure}

    \vspace{0.5cm} 

    \begin{subfigure}[b]{0.75\textwidth}
        \centering
        \includegraphics[width=\textwidth]{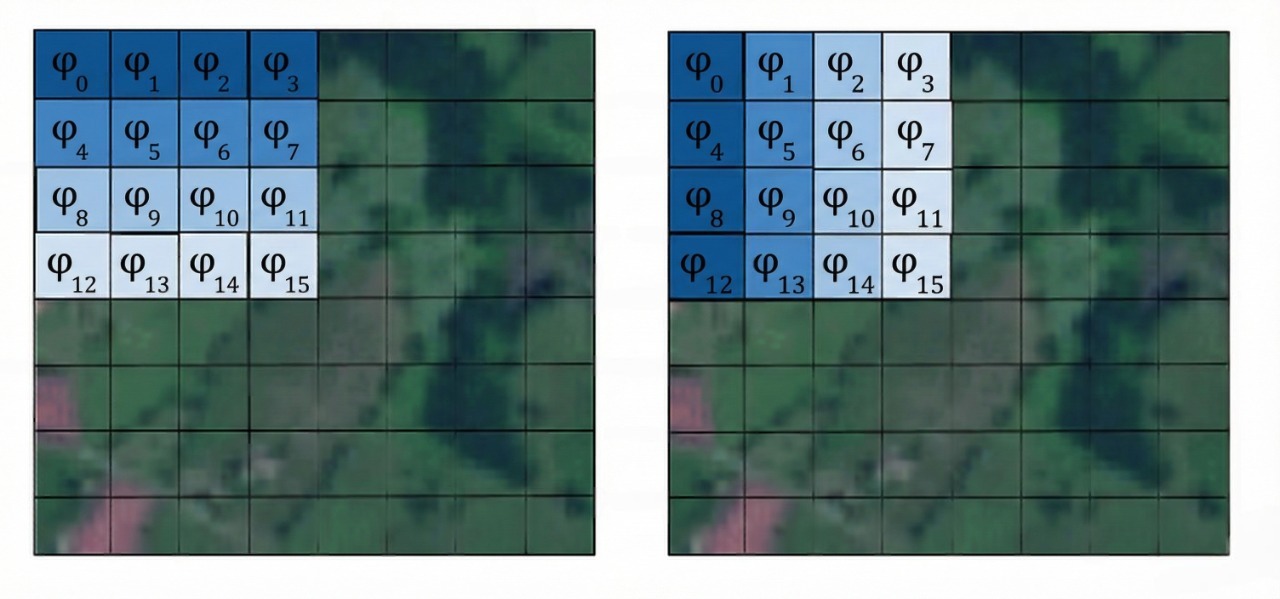}
        \caption{With this particular ordering we can mix both vertical and horizontal lines over the image.}
        \label{fig:v&h-lines}
    \end{subfigure}
    \caption{Diagram of how the pixels are ordered in the amplitudes of the state vector of the $N$-qubit circuit. For each ordering, the picture on the left (right) represents which pixels the first (second) gate is mixing. Pixels shaded with the same color are mixed together. }
    \label{fig:orderings}
\end{figure}

\newpage
\section{HQNN Parallel Hybrid model}

\begin{figure}[!htbp]
    \centering
    \begin{subfigure}{0.48\linewidth}
        \centering
        \includegraphics[width=\linewidth]{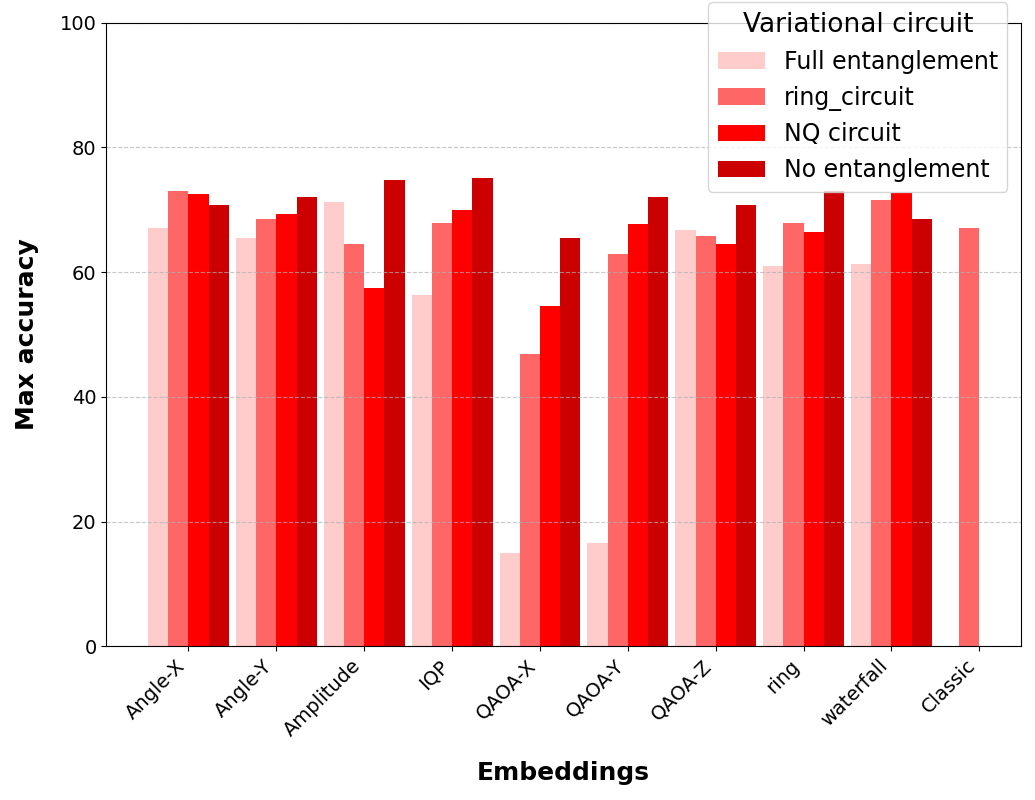}
        \caption{Best validation accuracies for $32\times 32$ images.}
        \label{parallel-val3232}
    \end{subfigure}
    \hfill
    \begin{subfigure}{0.48\linewidth}
        \centering
        \includegraphics[width=\linewidth]{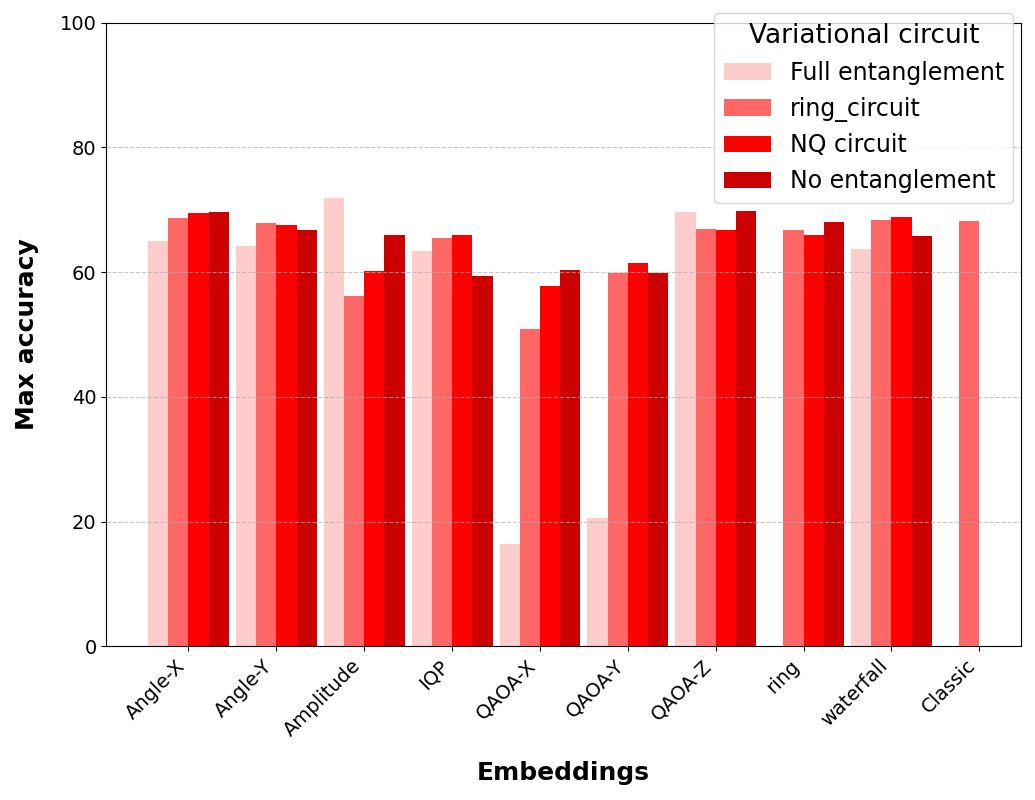}
        \caption{Best validation accuracies for $16\times 16$ images.}
        \label{parallel-val1616}
    \end{subfigure}

    \caption{Best validation accuracies with the HQNN-Parallel model. The result has been taken for the best combination encoding-ansätze and for the best measurement.}
    \label{parallel-best-acc}
\end{figure}

\begin{figure}[!htbp]
    \centering
    \begin{subfigure}{0.48\linewidth}
        \centering
        \includegraphics[width=\linewidth]{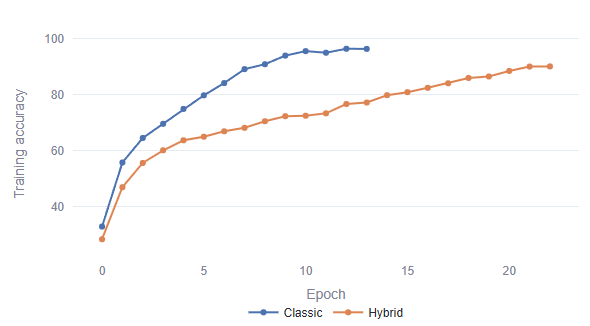}
        \caption{Train Accuracy.}
    \end{subfigure}
    \hfill
    \begin{subfigure}{0.48\linewidth}
        \centering
        \includegraphics[width=\linewidth]{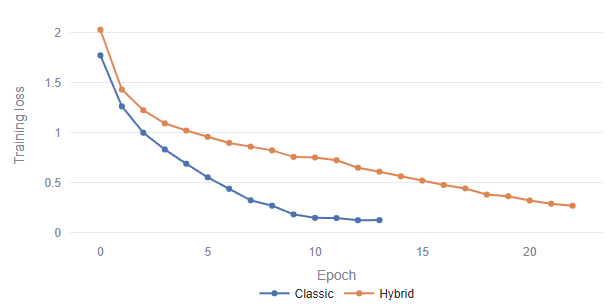}
        \caption{Train Loss.}
    \end{subfigure}

    \vspace{0.05\textwidth} 

    \begin{subfigure}{0.48\linewidth}
        \centering
        \includegraphics[width=\linewidth]{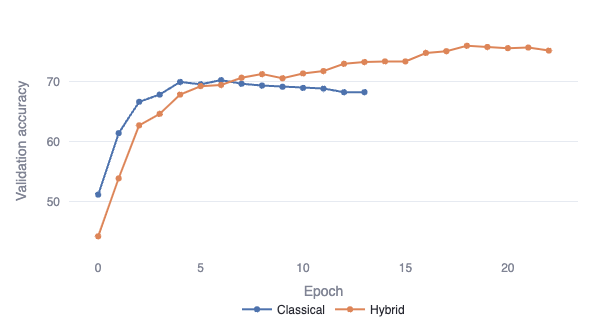}
        \caption{Validation Accuracy.}
    \end{subfigure}
    \hfill
    \begin{subfigure}{0.48\linewidth}
        \centering
        \includegraphics[width=\linewidth]{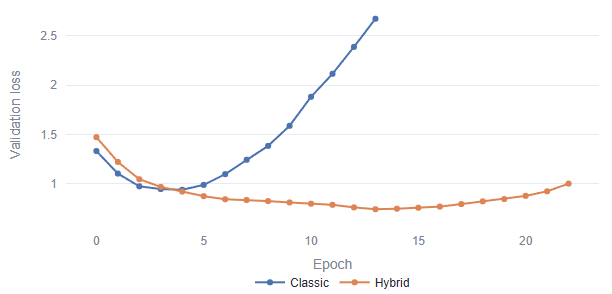}
        \caption{Validation Loss.}
    \end{subfigure}

    \caption{Training and validation accuracy and loss for the best HQNN-Parallel model (Amplitude-No entanglement) compared with the classical one.}
    \label{parallel-learning}
\end{figure}

\begin{figure}[!htbp]

\end{figure}

\begin{figure}[!htbp]
    \centering
    \begin{subfigure}{0.48\linewidth}
        \centering
     \includegraphics[width=\linewidth]{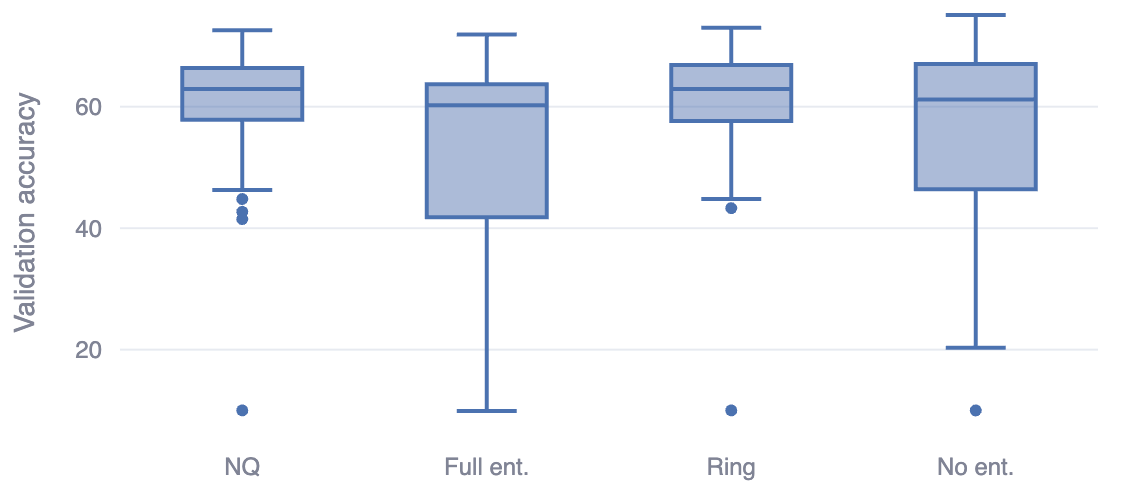}
     \subcaption{Validation accuracy of all configurations for a fixed ans\"atze.}
     \label{parallel-ansats-effect}
    \end{subfigure}
    \hfill
    \begin{subfigure}{0.48\linewidth}
        \centering
             \includegraphics[width=\linewidth]{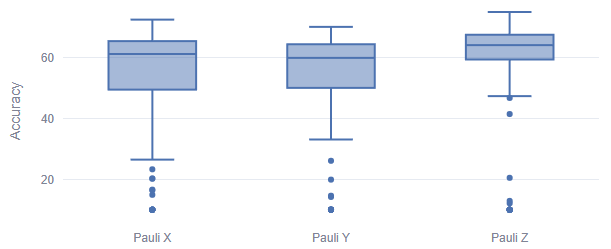}
         \subcaption{Validation accuracy of all configurations for a fixed measurement.}
              \label{parallel-meas-effect}
    \end{subfigure}
    \caption{Validation accuracies for different measurements and ans\"atze.}
    \label{parallel-measurement-effect}
\end{figure}

\begin{figure}[!htbp]
  \includegraphics[width=0.7\columnwidth]{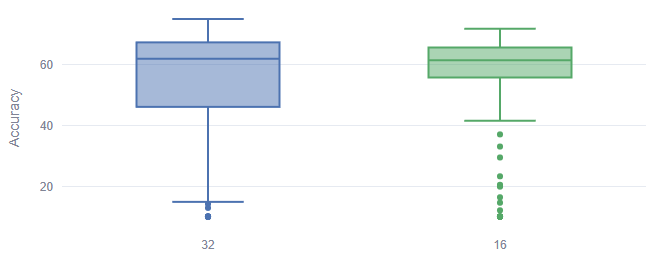}
  \caption{Validation accuracy across $32 \times 32$ and $16 \times 16$ image size.}
  \label{parallel-imgsize-effect}
\end{figure}

\break
\section{HQNN-Quanv model results}

\begin{figure}[!htbp]
    \centering
    \begin{subfigure}{0.48\linewidth}
        \centering
     \includegraphics[width=\linewidth]{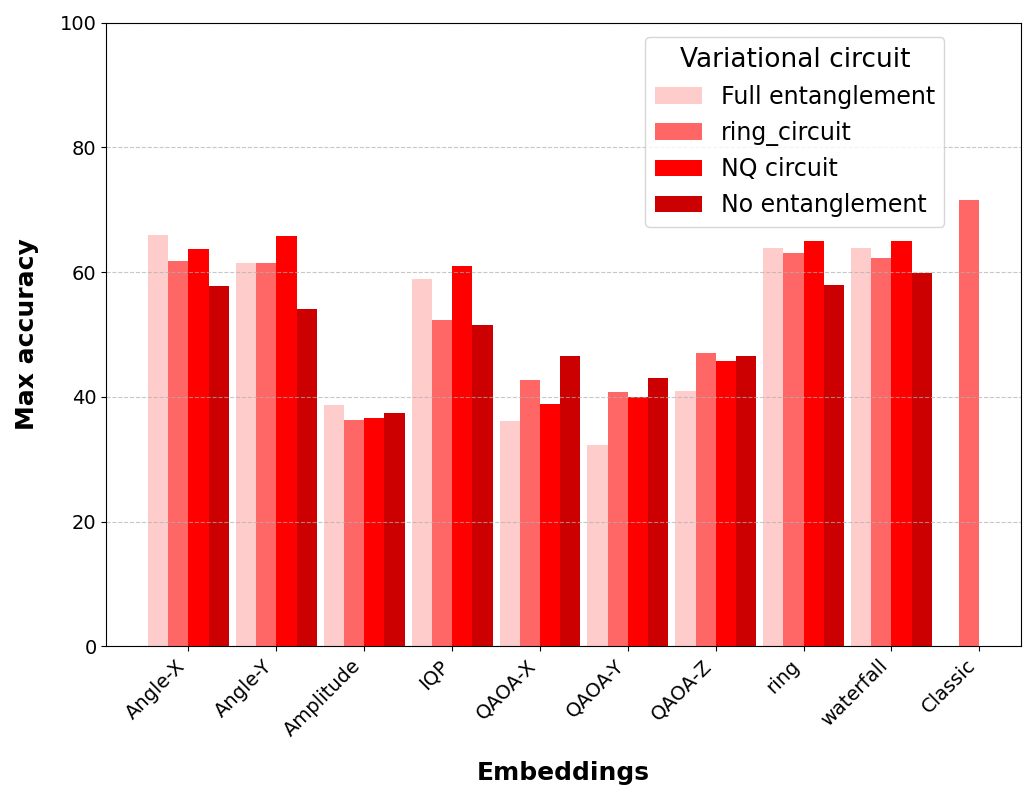}
     \subcaption{Best validation accuracies for $32\times 32$ images and qks = 3.}

    \end{subfigure}
    \hfill
    \begin{subfigure}{0.48\linewidth}
        \centering
     \includegraphics[width=\linewidth]{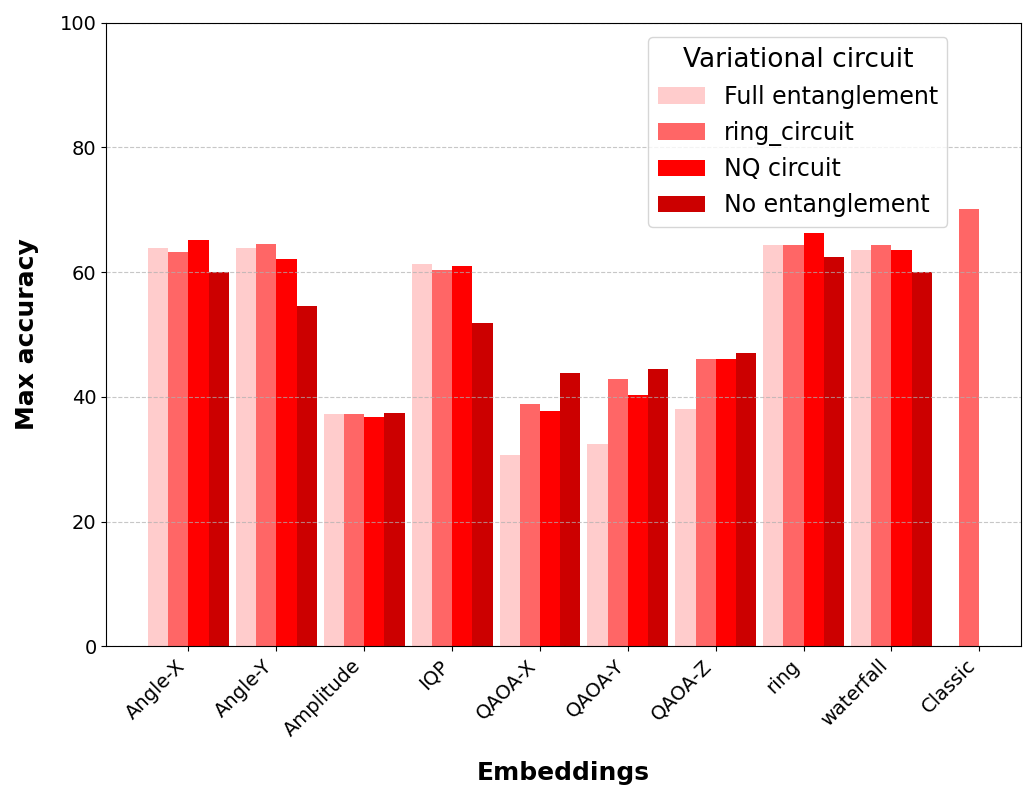}
     \subcaption{Best validation accuracies for $16\times 16$ images and qks = 3.}

    \end{subfigure}
    \caption{Best validation accuracies with the HQNN-Quanv model. The result has been taken for the best combination encoding-ans\"atze and for the best measurement.}
    \label{allflex}
\end{figure}

\begin{figure}[!htbp]
    \centering
    \begin{subfigure}{0.48\linewidth}
        \centering
        \includegraphics[width=\linewidth]{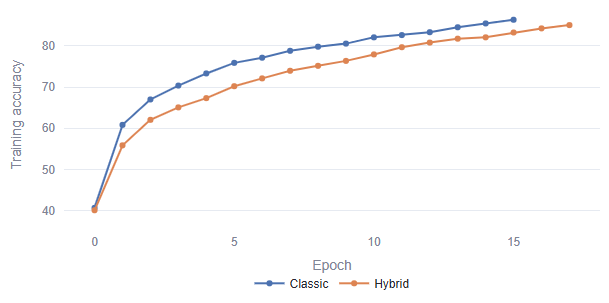}
        \caption{Train Accuracy.}
        \label{fig:flex_train_acc}
    \end{subfigure}
    \hfill
    \begin{subfigure}{0.48\linewidth}
        \centering
        \includegraphics[width=\linewidth]{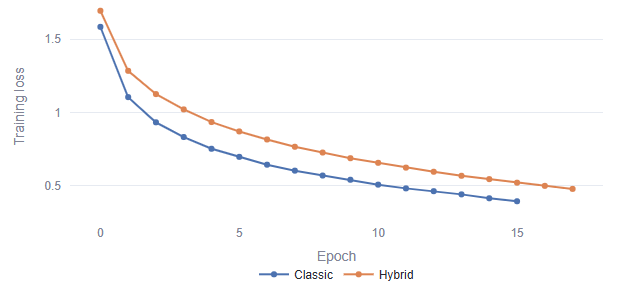}
        \caption{Train Loss.}
        \label{fig:flex_train_loss}
    \end{subfigure}

    \vspace{0.5cm}

    \begin{subfigure}{0.48\linewidth}
        \centering
        \includegraphics[width=\linewidth]{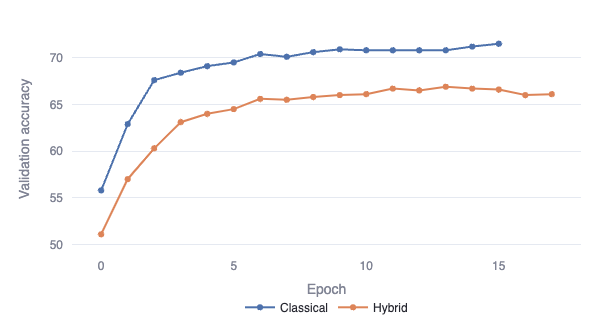}
        \caption{Validation Accuracy.}
        \label{fig:flex_val_acc}
    \end{subfigure}
    \hfill
    \begin{subfigure}{0.48\linewidth}
        \centering
        \includegraphics[width=\linewidth]{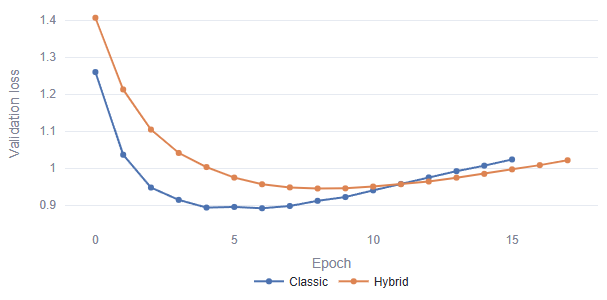}
        \caption{Validation Loss.}
        \label{fig:flex_val_loss}
    \end{subfigure}

    \caption{Train and validation accuracy and loss for the best HQNN-Quanv model (AngleX-Full entanglement) compared with the classical one.}
    \label{fig:flex_metrics_val}
\end{figure}

\begin{figure}[!htbp]
    \centering
    \begin{subfigure}{0.48\linewidth}
        \centering
        \includegraphics[width=\linewidth]{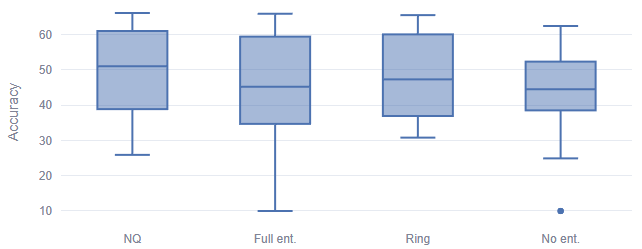}
        \subcaption{Ans\"atze.}
        \label{flex_ansatz}
    \end{subfigure}
    \hfill
    \begin{subfigure}{0.48\linewidth}
        \centering
        \includegraphics[width=\linewidth]{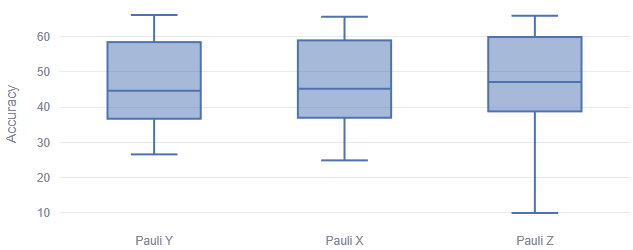}
        \subcaption{Measurements.}
        \label{flex_measurement}
    \end{subfigure}
    \caption{Validation accuracy comparison across ans\"atze and measurements. Image shapes ($32\times32$ and $16\times16$) and quantum kernel shapes (qks = 2, 3) are included.}
    \label{appendix_ans_measure}
\end{figure}

\begin{figure}[!htbp]
    \centering
    \begin{subfigure}{0.48\linewidth}
        \centering
        \includegraphics[width=\linewidth]{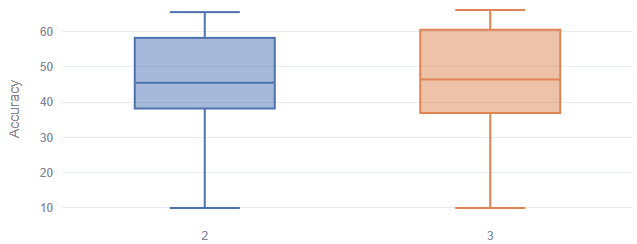}
        \subcaption{QKS shapes.}
        \label{flex_qks}
    \end{subfigure}
    \hfill
    \begin{subfigure}{0.48\linewidth}
        \centering
        \includegraphics[width=\linewidth]{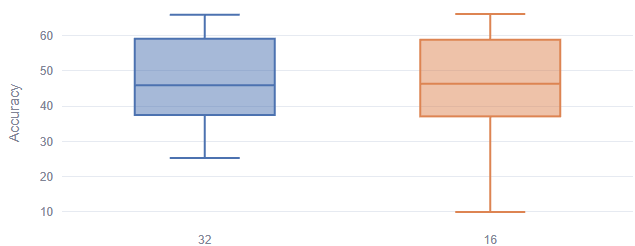}
        \subcaption{Image size: $32\times32$ and $16\times16$.}
        \label{flex_imgsize}
    \end{subfigure}
    \caption{Validation accuracy comparison across different QKS shapes and image size.}
    \label{appendix_qks_imgsize}
\end{figure}

\clearpage

\section{Pure quantum models results}

\begin{figure}[!htbp]
    \centering
    \begin{subfigure}{0.48\linewidth}
        \centering
        \includegraphics[width=\linewidth]{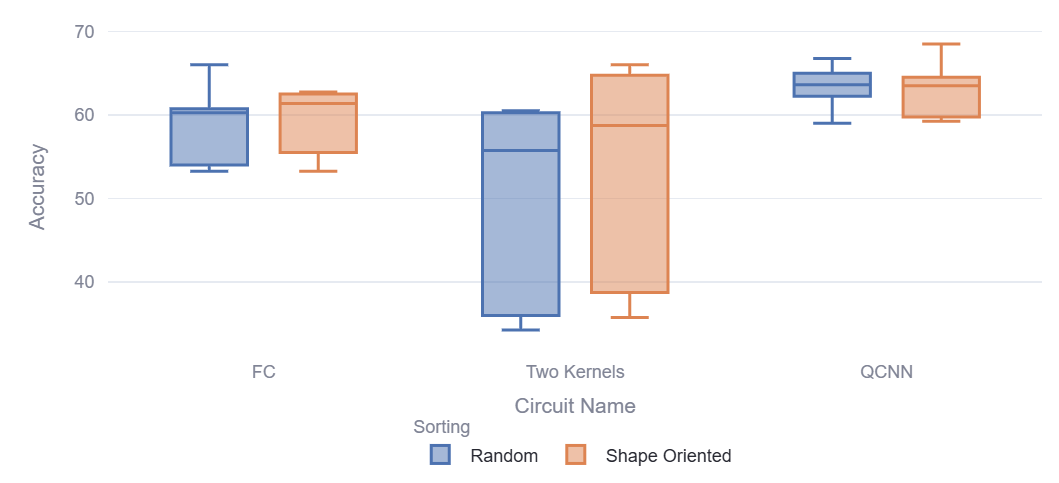}
        \subcaption{Orderings, 4 classes.}
        \label{qcnn_ord4}
    \end{subfigure}
    \hfill
    \begin{subfigure}{0.48\linewidth}
        \centering
        \includegraphics[width=\linewidth]{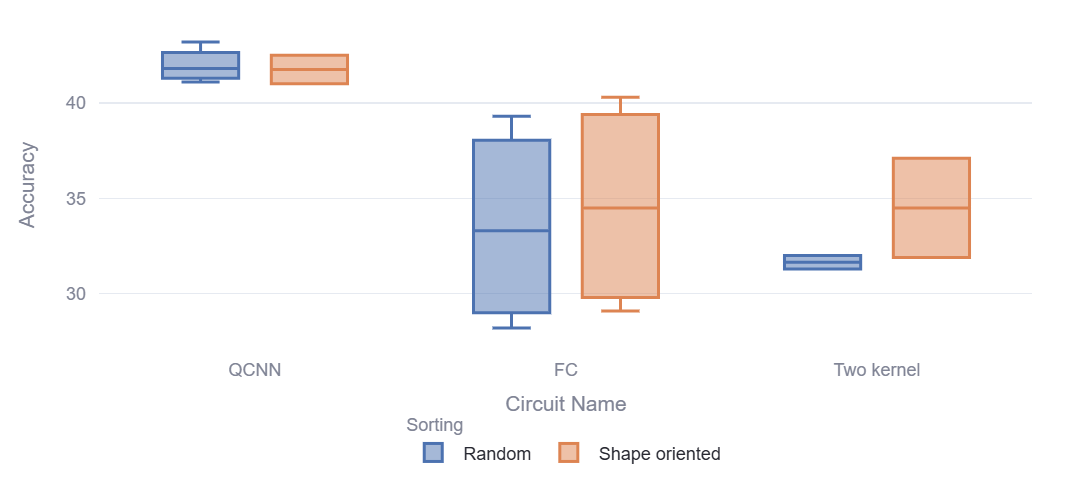}
        \subcaption{Orderings, 10 classes.}
        \label{qcnn_ord10}
    \end{subfigure}
    \caption{Validation accuracy comparison across different orderings for 4 and 10 classes separately.}
    \label{appendix_qcnn_orderings}
\end{figure}

\begin{figure}[!htbp]
    \centering
    \begin{subfigure}{0.48\linewidth}
        \centering
        \includegraphics[width=\linewidth]{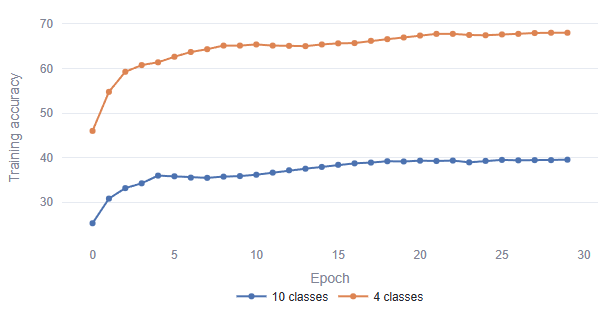}
        \caption{Train Accuracy.}
    \end{subfigure}
    \hfill
    \begin{subfigure}{0.48\linewidth}
        \centering
        \includegraphics[width=\linewidth]{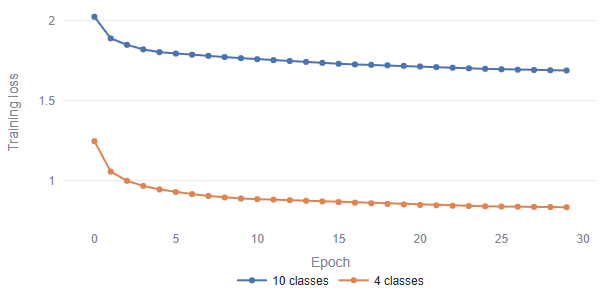}
        \caption{Train Loss.}
    \end{subfigure}

    \vspace{0.05\textwidth} 

    \begin{subfigure}{0.48\linewidth}
        \centering
        \includegraphics[width=\linewidth]{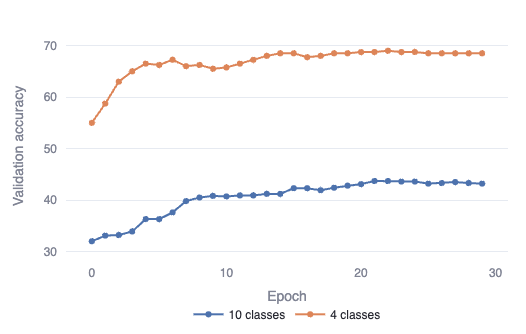}
        \caption{Validation Accuracy.}
    \end{subfigure}
    \hfill
    \begin{subfigure}{0.48\linewidth}
        \centering
        \includegraphics[width=\linewidth]{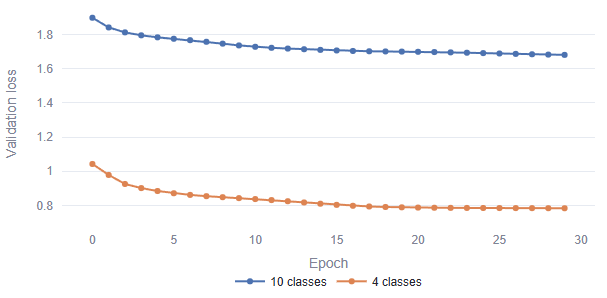}
        \caption{Validation Loss.}
    \end{subfigure}

    \caption{Training and validation accuracy and loss for the best purely quantum model for 4 (QCNN-Flatten-PauliZ) and 10 (QCNN-Flatten-Paulis) classes.}
    \label{qcnn_metrics}
\end{figure}

\begin{figure}[!htbp]
    \centering
    \begin{subfigure}{0.48\linewidth}
        \centering
        \includegraphics[width=\linewidth]{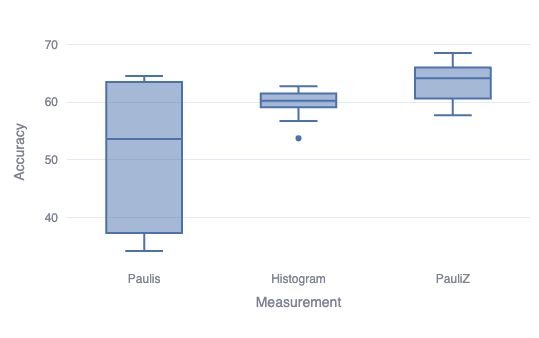}
        \subcaption{Measurements, 4 classes.}
        \label{qcnn_measurement4}
    \end{subfigure}
    \hfill
    \begin{subfigure}{0.48\linewidth}
        \centering
        \includegraphics[width=\linewidth]{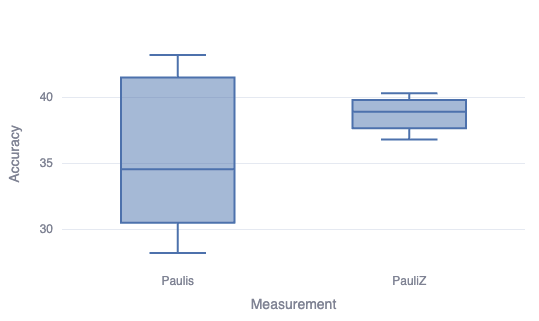}
        \subcaption{Measurements, 10 classes.}
        \label{qcnn_measurement10}
    \end{subfigure}
    \caption{Validation accuracy comparison across measurements for 4 and 10 classes separately.}
    \label{appendix_qcnn_measure}
\end{figure}

\begin{figure}[!htbp]
    \centering
    \begin{subfigure}{0.48\linewidth}
        \centering
        \includegraphics[width=\linewidth]{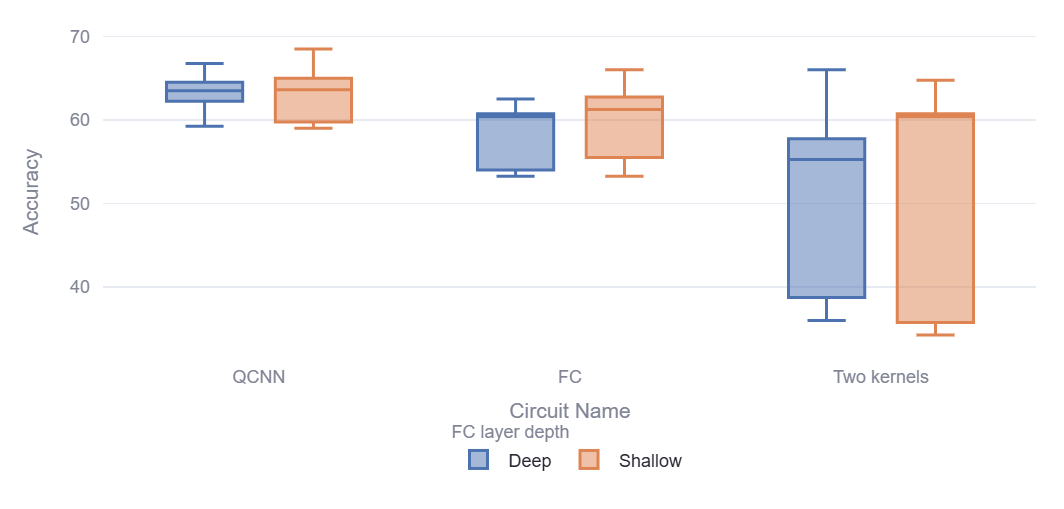}
        \subcaption{\emph{FC} layer depth, 4 classes.}
        \label{qcnn_fc_depth4}
    \end{subfigure}
    \hfill
    \begin{subfigure}{0.48\linewidth}
        \centering
        \includegraphics[width=\linewidth]{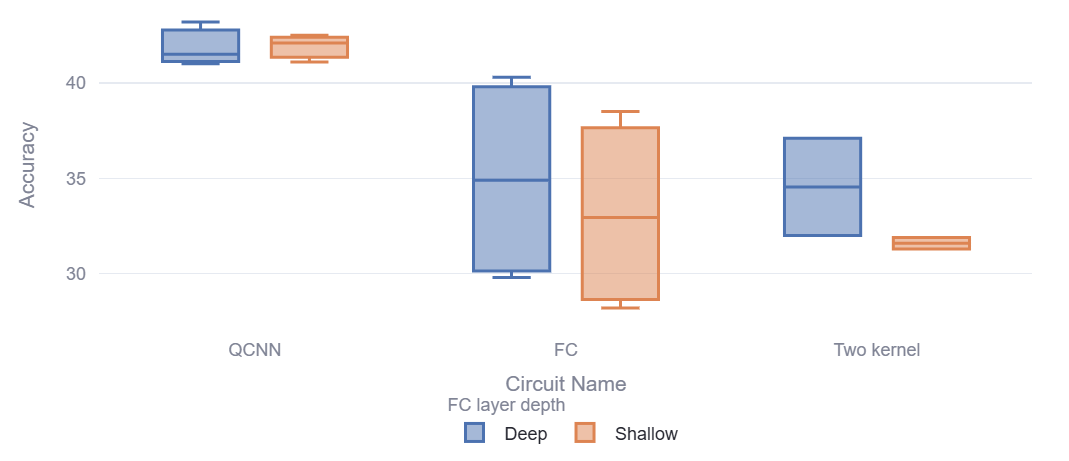}
        \subcaption{\emph{FC} layer depth, 10 classes.}
        \label{qcnn_fc_depth10}
    \end{subfigure}
    \caption{Validation accuracy comparison across \emph{FC} layer depth for 4 and 10 classes separately.}
    \label{appendix_qcnn_fc_layer_depth}
\end{figure}

\newpage
\section{QCNN and SEQNN, full results overview}

\begin{table}[h!]
\begin{tabular}{llllll}
\hline
\textbf{Ansatz}                             & \textbf{Ordering}        & \textbf{Measurement} & \textbf{4 classes} & \textbf{10 classes} & \textbf{N params}    \\ \hline
\multirow{6}{*}{QCNN}                       & \multirow{3}{*}{Random}  & histogram               & 59                 & -                   & 300                  \\
                                            &                          & PauliZ                & 65                 & -                   & 293                  \\
                                            &                          & Paulis               & 63.75              & 42.1                & 300                  \\ \cline{2-6} 
                                            & \multirow{3}{*}{Flatten} & histogram               & 59.75              & -                   & 300                  \\
                                            &                          & PauliZ                & 68.5               & -                   & 293                  \\
                                            &                          & Paulis               & 63.5               & 42.5                & 300                  \\ \hline
\multirow{6}{*}{Two kernel}                 & \multirow{3}{*}{Random}  & histogram               & 60.5               & -                   & \multirow{3}{*}{184} \\
                                            &                          & PauliZ                & 60.25              & -                   &                      \\
                                            &                          & Paulis               & 34.25              & 31.3                &                      \\ \cline{2-6} 
                                            & \multirow{3}{*}{Squared} & histogram               & 60.75              & -                   & \multirow{3}{*}{184} \\
                                            &                          & PauliZ                & 64.75              & -                   &                      \\
                                            &                          & Paulis               & 35.75              & 31.9                &                      \\ \hline
\multirow{6}{*}{FC (Simplified Two Design)} & \multirow{3}{*}{Random}  & histogram               & 60.25              & -                   & \multirow{3}{*}{190} \\
                                            &                          & PauliZ                & 66                 & 36.8                &                      \\
                                            &                          & Paulis               & 53.25              & 28.2                &                      \\ \cline{2-6} 
                                            & \multirow{3}{*}{Squared} & histogram               & 62.75              & -                   & \multirow{3}{*}{190} \\
                                            &                          & PauliZ                & 62.25              & 38.5                &                      \\
                                            &                          & Paulis               & 55.5               & 29.1                &                      \\ \hline
\end{tabular}
\caption{Validation accuracy results for shallow FC layer.}
\end{table}

\begin{figure}[!htbp]
    \centering
    \begin{subfigure}{0.48\linewidth}
        \centering
        \includegraphics[width=\linewidth]{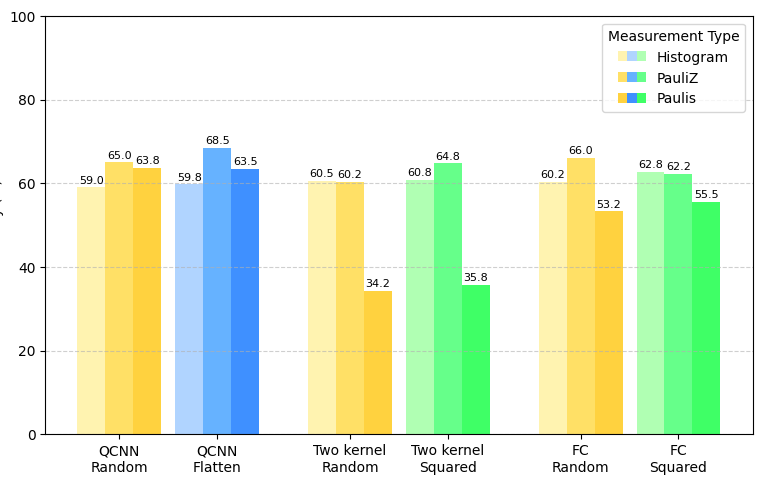}
    \end{subfigure}
    \hfill
    \begin{subfigure}{0.48\linewidth}
        \centering
        \includegraphics[width=\linewidth]{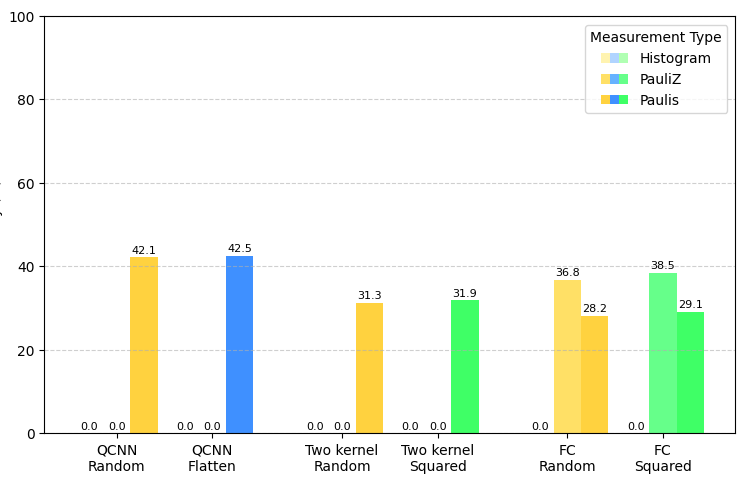}
    \end{subfigure}
    \caption{Validation accuracy by Ansatz, Ordering, and Measurement type for 4 and 10 classes respectively. Shallow FC layer.}
    \label{qcnnaccuracies}
\end{figure}

\begin{table}[h!]
\begin{tabular}{llllll}
\hline
\textbf{Ansatz}                               & \textbf{Ordering}        & \textbf{Measurement} & \textbf{4 classes} & \textbf{10 classes} & \textbf{N params}    \\ \hline
\multirow{6}{*}{QCNN}                         & \multirow{3}{*}{Random}  & histogram               & 62.25              & -                   & 326                  \\
                                              &                          & PauliZ                & 66.75              & -                   & 373                  \\
                                              &                          & Paulis               & 63.5               & 41.5                & 316                  \\ \cline{2-6} 
                                              & \multirow{3}{*}{Flatten} & histogram               & 59.25              & -                   & 326                  \\
                                              &                          & PauliZ                & 63.5               & -                   & 373                  \\
                                              &                          & Paulis               & 64.5               & 41                  & 316                  \\ \hline
\multirow{6}{*}{Two kernel}                   & \multirow{3}{*}{Random}  & histogram               & 53.75              & -                   & \multirow{3}{*}{310} \\
                                              &                          & PauliZ                & 57.75              & -                   &                      \\
                                              &                          & Paulis               & 36                 & 32                  &                      \\ \cline{2-6}
                                              & \multirow{3}{*}{Squared} & histogram               & 56.75              & -                   & \multirow{3}{*}{310} \\
                                              &                          & PauliZ                & 66                 & -                   &                      \\
                                              &                          & Paulis               & 38.75              & 37.1                &                      \\ \hline
\multirow{6}{*}{FC (Simplified Two   Design)} & \multirow{3}{*}{Random}  & histogram               & 60.25              & -                   & \multirow{3}{*}{352} \\
                                              &                          & PauliZ                & 60.75              & 39.3                &                      \\
                                              &                          & Paulis               & 54                 & 29.8                &                      \\ \cline{2-6}
                                              & \multirow{3}{*}{Squared} & histogram               & 62.5               & -                   & \multirow{3}{*}{352} \\
                                              &                          & PauliZ                & 60.5               & 40.3                &                      \\
                                              &                          & Paulis               & 53.25              & 30.5                &                      \\ \hline
\end{tabular}
\caption{Results for deep FC layer.}
\end{table}

\begin{figure}[!htbp]
    \centering
    \begin{subfigure}{0.48\linewidth}
        \centering
        \includegraphics[width=\linewidth]{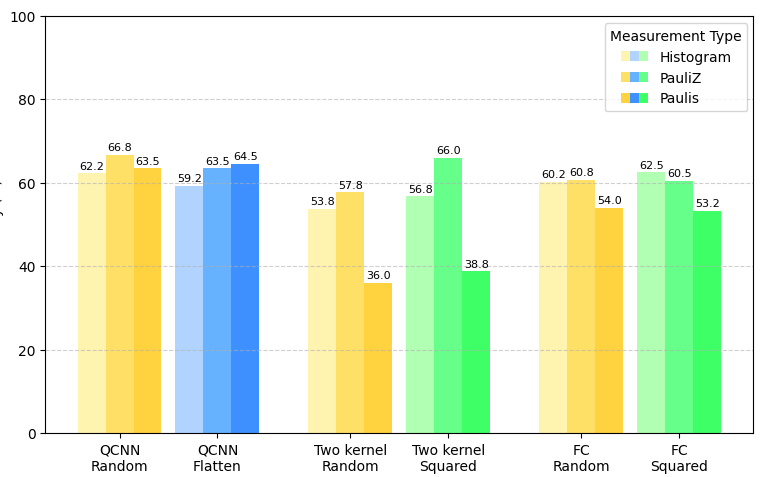}
    \end{subfigure}
    \hfill
    \begin{subfigure}{0.48\linewidth}
        \centering
        \includegraphics[width=\linewidth]{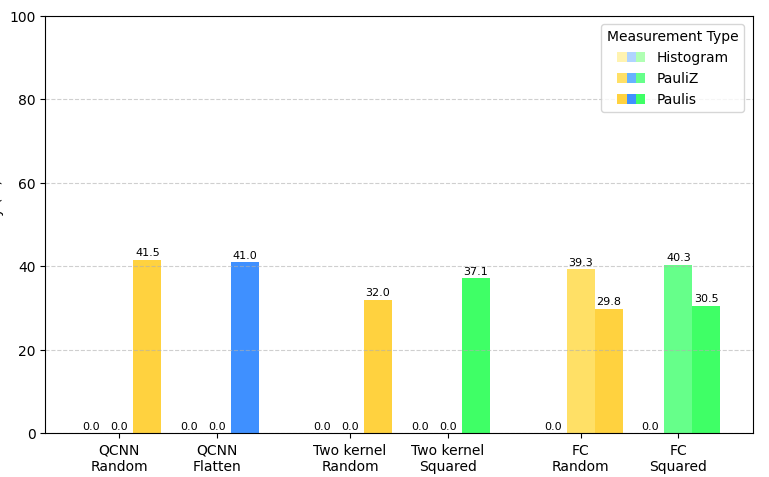}
    \end{subfigure}
    \caption{Validation accuracy by Ansatz, Ordering, and Measurement type for 4 and 10 classes respectively. Deep FC layer.}
    \label{qcnnaccuracies2}
\end{figure}

\clearpage

\section{Expressibility of QAOA encoding}\label{appendix:expressibility}

In this section we calculate an estimate of the \emph{state frame potential} for the $QAOA$ encoding for the three possible local fields. The \emph{state frame potential} has been used before as a measure of expressibility by comparing with that of the ensemble of Haar distributed states \cite{sim2019expressibility}. We compare the first two moments, $\mathcal{F}^{(1)}$ and $\mathcal{F}^{(2)}$ of the frame potential normalized by the value for Haar states for the three local fields. For that, we calculate the measure over several fixed random inputs. The lower the measure, the closer the distribution of states generated by the encoding circuits is to the Haar distribution; hence, more expressive. Results are conveyed in \hyperref[table-expressibility-4]{TABLE \ref{table-expressibility-4}} and \hyperref[table-expressibility-8]{TABLE \ref{table-expressibility-8}}.

\begin{table}[ht!]
\begin{tabular}{lll}
\hline
\textbf{Local field} & $\mathbf{\mathcal{F}^{(1)}/\mathcal{F}^{(1)}_{Haar}}$ & $\mathbf{\mathcal{F}^{(2)}/\mathcal{F}^{(2)}_{Haar}}$ \\ \hline
X                    & $10.02(0.23)$           & $55.6(2.5)$          \\
Y                    & $9.67(0.23)$           & $52.6(2.3)$           \\
Z                    & $12.5(1.0)$           & $96(14)$           \\ \hline
\end{tabular}
\caption{Results for the \emph{state frame potential} for $4$ qubits.}
\label{table-expressibility-4}
\end{table}

\begin{table}[ht!]
\begin{tabular}{lll}
\hline
\textbf{Local field} & $\mathbf{\mathcal{F}^{(1)}/\mathcal{F}^{(1)}_{Haar}}$ & $\mathbf{\mathcal{F}^{(2)}/\mathcal{F}^{(2)}_{Haar}}$ \\ \hline
X                    & $100.0(4.6)$           & $5550 (460)$          \\
Y                    & $96.1(3.4)$           & $5220(360)$           \\
Z                    & $154(21)$           & $12800(3400)$           \\ \hline
\end{tabular}
\caption{Results for the \emph{state frame potential} for $8$ qubits.}
\label{table-expressibility-8}
\end{table}

\end{document}